\documentclass[11pt]{article}
\usepackage[dvips]{graphicx}
\usepackage{float}
\usepackage{epsfig}
\usepackage{ulem}
\usepackage{latexsym,amsmath,amsfonts,amssymb}
\usepackage[latin1]{inputenc}
\usepackage{rotating}
\usepackage[american]{babel}
\usepackage[dvips]{graphicx}
\usepackage{bbm}
\usepackage{color}
\usepackage{slashed}
\usepackage[unicode]{hyperref}
\usepackage{lscape}
\usepackage{bigints}
\usepackage{enumerate}
\usepackage[shortlabels]{enumitem}
\usepackage{tikz}
\usetikzlibrary{decorations.pathreplacing}
\usetikzlibrary{shapes}
\usepackage{graphicx}
\usepackage{epsfig}
\usepackage{rotating}
\usepackage{amssymb}
\usepackage{subfigure}
\usepackage{dsfont}
\usepackage{psfrag}
\usepackage{amsmath,euscript,array,mathrsfs,amsfonts}
\usepackage{slashed}
\usepackage{array}
\usepackage{youngtab}
\usepackage{color}
\usepackage{bbold}
\usepackage{hyperref}
\hypersetup{
colorlinks = true,
linkcolor = red,
linktocpage = true,
citecolor = blue
}

\usepackage{tikz}
\usetikzlibrary{calc}
 \usetikzlibrary{decorations.text}
 \usetikzlibrary{shapes}

 \usetikzlibrary{decorations.pathmorphing}
\usetikzlibrary{decorations.pathreplacing}
\usetikzlibrary{arrows.meta}
\tikzset{
  >={To[length=5pt]}
  }
\usetikzlibrary{shapes, shapes.geometric, shapes.symbols, shapes.arrows, shapes.multipart, shapes.callouts, shapes.misc}
\tikzset{snake it/.style={decorate, decoration=snake}}
\tikzset{7brane/.style={circle, draw=black, fill=black,ultra thick,inner sep=1.5 pt, minimum size=1 pt,}, c/.default={4pt}}
\tikzset{cross/.style={cross out, draw=black,thick, minimum size=2*(#1-\pgflinewidth), inner sep=0pt, outer sep=0pt}, cross/.default={5pt}}
\tikzset{big7brane/.style={circle, draw=black, fill=black,ultra thick,inner sep=2.5 pt, minimum size=1 pt,}, c/.default={4pt}}
\tikzset{u/.style={circle, draw=black, fill=white,inner sep=2 pt, minimum size=2 pt,},f/.style={square, draw=black, fill=white,ultra thick,inner sep=4 pt, minimum size=2 pt,}}
\tikzset{so/.style={circle, draw=black, fill=red,inner sep=2 pt, minimum size=2 pt,},f/.style={square, draw=black, fill=white,ultra thick,inner sep=4 pt, minimum size=2 pt,}}
\tikzset{sp/.style={circle, draw=black, fill=blue,inner sep=2 pt, minimum size=2 pt,},f/.style={square, draw=black, fill=white,ultra thick,inner sep=4 pt, minimum size=2 pt,}}
\tikzset{uf/.style={rectangle, draw=black, fill=white,inner sep=3 pt, minimum size=4 pt,}}
\tikzset{spf/.style={rectangle, draw=black, fill=blue, thick,inner sep=3 pt, minimum size=4 pt, circle, draw=black, fill=blue,thick,inner sep=2 pt, minimum size=2 pt,},f/.style={square, draw=black, fill=white,ultra thick,inner sep=4 pt, minimum size=2 pt,}}
\tikzset{sof/.style={rectangle, draw=black, fill=red, thick,inner sep=3 pt, minimum size=4 pt,}}
\usetikzlibrary{positioning}
\usetikzlibrary{arrows}
\usetikzlibrary{decorations.pathreplacing}
\usetikzlibrary{shapes}

\makeatletter\def\l@subsubsection#1#2{}%
\makeatother

\renewcommand\theequation{\arabic{section}.\arabic{equation}} 

\usepackage{subfigure}

\def\cA{{\cal A}}

\def\cF{{\cal F}}

\def\CC{\ensuremath{\mathds C}}
\def\RR{\ensuremath{\mathds R}}
\def\ZZ{\ensuremath{\mathds Z}}

\DeclareMathOperator{\vol}{vol}

\DeclareMathOperator{\sech}{sech}
\DeclareMathOperator{\tr}{tr}

\DeclareMathOperator{\csch}{csch}

\newcommand{\be}{\begin{equation}}
\newcommand{\ee}{\end{equation}}
\newcommand{\ba}{\begin{array}}
\newcommand{\ea}{\end{array}}

\def\im{Invent. Math.}

\def\hat{\widehat}
\def\a{\alpha}
\def\b{\beta}
\def\c{\gamma}
\def\d{\delta}
\def\f{\phi}               
\def\vf{\varphi}  
\def\tvf{\tilde{\varphi}}
\def\vp{\varphi}
\def\g{\gamma}
\def\h{\eta}
\def\j{\psi}
\def\k{\kappa}                    
\def\l{\lambda}
\def\m{\mu}
\def\n{\nu}
\def\o{\omega}  \def\w{\omega}
\def\p{\pi}
\def\q{\theta}  \def\th{\theta}                  
\def\r{\rho}                                     
\def\s{\sigma}                                   
\def\t{\tau}
\def\u{\upsilon}
\def\x{\xi}
\def\z{\zeta}
\def\pt{\tilde{\varphi}}
\def\tt{\tilde{\theta}}
\def\lab{\label}
\def\6{\partial}
\def\wg{\wedge}
\def\bpsi{\bar{\psi}}
\def\bt{\bar{\theta}}
\def\bvf{\bar{\varphi}}
\def\W{\Omega}

\newcommand{\td}{\mathrm{d}}

\DeclareMathOperator{\str}{str}

\newcommand{\beq}{\begin{equation}}
\newcommand{\eeq}{\end{equation}}
\newcommand{\bea}{\begin{eqnarray}}
\newcommand{\eea}{\end{eqnarray}}
\newcommand{\nn}{\nonumber}

\newcommand{\beqs}{\begin{eqnarray}}
\newcommand{\eeqs}{\end{eqnarray}}
\newcommand{\bal}{\begin{aligned}}
\newcommand{\eal}{\end{aligned}}
\makeatletter
\newcommand\setItemnumber[1]{\setcounter{enum\romannumeral\@enumdepth}{\numexpr#1-1\relax}}
\makeatother
\begin{document}
\baselineskip=15.5pt
\pagestyle{plain}
\setcounter{page}{1}

\def\del{{\partial}}
\def\vev#1{\left\langle #1 \right\rangle}
\def\cn{{\cal N}}
\def\co{{\cal O}}


\def\IC{{\mathbb C}}
\def\IR{{\mathbb R}}
\def\IZ{{\mathbb Z}}
\def\RP{{\bf RP}}
\def\CP{{\bf CP}}
\def\Poincaré{{Poincar\'e }}
\def\tr{{\rm tr}}
\def\tp{{\tilde \Phi}}

\def\TL{\hfil$\displaystyle{##}$}
\def\TR{$\displaystyle{{}##}$\hfil}
\def\TC{\hfil$\displaystyle{##}$\hfil}
\def\TT{\hbox{##}}
\def\HLINE{\noalign{\vskip1\jot}\hline\noalign{\vskip1\jot}}
\def\seqalign#1#2{\vcenter{\openup1\jot
   \halign{\strut #1\cr #2 \cr}}}
\def\lbldef#1#2{\expandafter\gdef\csname #1\endcsname {#2}}
\def\eqn#1#2{\lbldef{#1}{(\ref{#1})}%
\begin{equation} #2 \label{#1} \end{equation}}
\def\eqalign#1{\vcenter{\openup1\jot
     \halign{\strut\span\TL & \span\TR\cr #1 \cr
    }}}

\def\eno#1{(\ref{#1})}
\def\href#1#2{#2}
\def\half{\frac{1}{2}}



\def\ads{{\it AdS}}
\def\adsp{{\it AdS}$_{p+2}$}
\def\cft{{\it CFT}}

\newcommand{\ber}{\begin{eqnarray}}
\newcommand{\eer}{\end{eqnarray}}

\newcommand{\beqar}{\begin{eqnarray}}
\newcommand{\cO}{{\cal O}}
\newcommand{\cT}{{\cal T}}
\newcommand{\cR}{{\cal R}}
\newcommand{\eeqar}{\end{eqnarray}}
\newcommand{\tht}{\thteta}
\newcommand{\lm}{\lambda}\newcommand{\Lm}{\Lambda}


\newcommand{\nonu}{\nonumber}
\newcommand{\oh}{\displaystyle{\frac{1}{2}}}
\newcommand{\dsl}
   {\kern.06em\hbox{\raise.15ex\hbox{$/$}\kern-.56em\hbox{$\partial$}}}
\newcommand{\as}{\not\!\! A}
\newcommand{\ps}{\not\! p}
\newcommand{\ks}{\not\! k}
\newcommand{\D}{{\cal{D}}}
\newcommand{\dv}{d^2x}
\newcommand{\Z}{{\cal Z}}
\newcommand{\N}{{\cal N}}
\newcommand{\Dsl}{\not\!\! D}
\newcommand{\Bsl}{\not\!\! B}
\newcommand{\Psl}{\not\!\! P}

\newcommand{\eeqarr}{\end{eqnarray}}


\def\del{{\delta^{\hbox{\sevenrm B}}}} \def\ex{{\hbox{\rm e}}}
\def\azb{A_{\bar z}} \def\az{A_z} \def\bzb{B_{\bar z}} \def\bz{B_z}
\def\czb{C_{\bar z}} \def\cz{C_z} \def\dzb{D_{\bar z}} \def\dz{D_z}
\def\im{{\hbox{\rm Im}}} \def\mod{{\hbox{\rm mod}}} \def\tr{{\hbox{\rm Tr}}}
\def\ch{{\hbox{\rm ch}}} \def\imp{{\hbox{\sevenrm Im}}}
\def\trp{{\hbox{\sevenrm Tr}}} \def\vol{{\hbox{\rm Vol}}}
\def\rl{\Lambda_{\hbox{\sevenrm R}}} \def\wl{\Lambda_{\hbox{\sevenrm W}}}
\def\fc{{\cal F}_{k+\cox}} \def\vev{vacuum expectation value}
\def\nodiv{\mid{\hbox{\hskip-7.8pt/}}}
\def\ie{{\em i.e.}}
\def\ie{\hbox{\it i.e.}}

\def\CC{{\mathchoice
{\rm C\mkern-8mu\vrule height1.45ex depth-.05ex
width.05em\mkern9mu\kern-.05em}
{\rm C\mkern-8mu\vrule height1.45ex depth-.05ex
width.05em\mkern9mu\kern-.05em}
{\rm C\mkern-8mu\vrule height1ex depth-.07ex
width.035em\mkern9mu\kern-.035em}
{\rm C\mkern-8mu\vrule height.65ex depth-.1ex
width.025em\mkern8mu\kern-.025em}}}

\def\RR{{\rm I\kern-1.6pt {\rm R}}}
\def\NN{{\rm I\!N}}
\def\ZZ{{\rm Z}\kern-3.8pt {\rm Z} \kern2pt}
\def\IB{\relax{\rm I\kern-.18em B}}
\def\ID{\relax{\rm I\kern-.18em D}}
\def\II{\relax{\rm I\kern-.18em I}}
\def\IP{\relax{\rm I\kern-.18em P}}
\newcommand{\CS}{{\scriptstyle {\rm CS}}}
\newcommand{\CSs}{{\scriptscriptstyle {\rm CS}}}
\newcommand{\rc}{\nonumber\\}
\newcommand{\bear}{\begin{eqnarray}}
\newcommand{\eear}{\end{eqnarray}}

\newcommand{\LL}{{\cal L}}

\def\mani{{\cal M}}
\def\calo{{\cal O}}
\def\calb{{\cal B}}
\def\calw{{\cal W}}
\def\calz{{\cal Z}}
\def\cald{{\cal D}}
\def\calc{{\cal C}}
\newcommand{\gt}{\tilde{g}}

\def\to{\rightarrow}
\def\ele{{\hbox{\sevenrm L}}}
\def\ere{{\hbox{\sevenrm R}}}
\def\zb{{\bar z}}
\def\wb{{\bar w}}
\def\nodiv{\mid{\hbox{\hskip-7.8pt/}}}
\def\menos{\hbox{\hskip-2.9pt}}
\def\dr{\dot R_}
\def\drr{\dot r_}
\def\ds{\dot s_}
\def\da{\dot A_}
\def\dga{\dot \gamma_}
\def\ga{\gamma_}
\def\dal{\dot\alpha_}
\def\al{\alpha_}
\def\cl{{closed}}
\def\cls{{closing}}
\def\vev{vacuum expectation value}
\def\tr{{\rm Tr}}
\def\to{\rightarrow}
\def\too{\longrightarrow}


\def\a{\alpha}
\def\b{\beta}
\def\c{\gamma}
\def\d{\delta}
\def\e{\epsilon}           
\def\F{\Phi}
\def\f{\phi}               
\def\vf{\varphi}  \def\tvf{\tilde{\varphi}}
\def\vp{\varphi}
\def\g{\gamma}
\def\h{\eta}
\def\j{\psi}
\def\k{\kappa}                    
\def\l{\lambda}
\def\m{\mu}
\def\n{\nu}
\def\o{\omega}  \def\w{\omega}
\def\q{\theta}  \def\th{\theta}                  
\def\r{\rho}                                     
\def\s{\sigma}                                   
\def\t{\tau}
\def\u{\upsilon}
\def\x{\xi}
\def\X{\Xi}
\def\z{\zeta}
\def\pt{\tilde{\varphi}}
\def\tt{\tilde{\theta}}
\def\lab{\label}
\def\6{\partial}
\def\wg{\wedge}
\def\atanh{{\rm arctanh}}
\def\bpsi{\bar{\psi}}
\def\bt{\bar{\theta}}
\def\bvf{\bar{\varphi}}

\def\ft#1#2{{\textstyle{{\scriptstyle #1}\over {\scriptstyle #2}}}}
\def\fft#1#2{{#1 \over #2}}
\def\del{\partial}
\def\sst#1{{\scriptscriptstyle #1}}

\def\dalemb#1#2{{\vbox{\hrule height .#2pt
        \hbox{\vrule width.#2pt height#1pt \kern#1pt
                \vrule width.#2pt}
        \hrule height.#2pt}}}
\def\square{\mathord{\dalemb{6.8}{7}\hbox{\hskip1pt}}}
\def\hF{\hat F}
\def\tA{\widetilde A}
\def\tcA{{\widetilde{\cal A}}}
\def\tcF{{\widetilde{\cal F}}}
\def\hA{\hat{\cal A}}
\def\cF{{\cal F}}
\def\cA{{\cal A}}
\def\wdg{{\sst \wedge}}

\def\0{{\sst{(0)}}}
\def\1{{\sst{(1)}}}
\def\2{{\sst{(2)}}}
\def\3{{\sst{(3)}}}
\def\4{{\sst{(4)}}}
\def\5{{\sst{(5)}}}
\def\6{{\sst{(6)}}}
\def\7{{\sst{(7)}}}
\def\8{{\sst{(8)}}}
\def\n{{\sst{(n)}}}
\def\tV{\widetilde V}
\def\tW{\widetilde W}
\def\tH{\widetilde H}
\def\tE{\widetilde E}
\def\tF{\widetilde F}
\def\tA{\widetilde A}
\def\tP{{\widetilde P}}
\def\tD{\widetilde D}
\def\bA{\bar{\cal A}}
\def\bF{\bar{\cal F}}
\def\tG{\widetilde G}
\def\tT{\widetilde T}
\def\Z{\rlap{\sf Z}\mkern3mu{\sf Z}}
\def\R{\rlap{\rm I}\mkern3mu{\rm R}}
\def\G{{\cal G}}
\def\gg{\bf g}
\def\CS{{\cal S}}
\def\S{{\cal S}}
\def\P{{\cal P}}
\def\ep{\epsilon}
\def\td{\tilde}
\def\wtd{\widetilde}
\def\half{{\textstyle{1\over2}}}
\def\Qw{{Q_{\rm wave}}}
\def\Qnut{{Q_{\sst{\rm NUT}}}}
\def\mun{{\mu_{\sst{\rm NUT}}}}
\def\muw{{\mu_{\rm wave}}}
\let\a=\alpha \let\b=\beta \let\g=\gamma \let\d=\delta \let\e=\epsilon
\let\z=\zeta \let\h=\eta \let\q=\theta \let\i=\iota \let\k=\kappa
\let\l=\lambda \let\m=\mu \let\n=\nu \let\x=\xi \let\p=\pi \let\r=\rho
\let\s=\sigma \let\t=\tau \let\u=\upsilon \let\f=\phi \let\c=\chi \let\y=\psi
\let\w=\omega  \let\D=\Delta \let\Q=\Theta \let\L=\Lambda
\let\X=\Xi  \let\U=\Upsilon \let\F=\Phi \let\Y=\Psi
\let\C=\Chi \let\W=\Omega     
\let\la=\label \let\ci=\cite \let\re=\ref
\let\se=\section \let\sse=\subsection \let\ssse=\subsubsection 
\def\nn{\nonumber} \def\bd{\begin{document}} \def\ed{\end{document}}
\def\ds{\documentstyle} \let\fr=\frac \let\bl=\bigl \let\br=\bigr
\let\Br=\Bigr \let\Bl=\Bigl 
\let\bm=\bibitem
\let\na=\nabla
\let\pa=\partial \let\ov=\overline 
\def\ba{\begin{eqnarray}}
\def\ea{\end{eqnarray}}
\def\ft#1#2{{\textstyle{{\scriptstyle #1}\over {\scriptstyle #2}}}}
\def\fft#1#2{{#1 \over #2}}
\def\del{\partial}
\def\sst#1{{\scriptscriptstyle #1}}
\def\oneone{\rlap 1\mkern4mu{\rm l}}
\def\ie{{\it i.e.\ }}
\def\via{{\it via}}
\def\semi{{\ltimes}}
\def\str{{\rm str}}
\def\jm{{\rm j}}
\def\im{{\rm i}}
\def\mapright#1{\smash{\mathop{-\!\!\!-\!\!\!-\!\!\!-\!\!\!-\!\!\!
             \longrightarrow}\limits^{#1}}}
\def\maprightt#1#2{\smash{\mathop{-\!\!\!-\!\!\!-\!\!\!-\!\!\!-\!\!\!
             \longrightarrow}\limits^{#1}_{#2}}}

\newcommand{\ho}[1]{$\, ^{#1}$}
\newcommand{\hoch}[1]{$\, ^{#1}$}
\newcommand{\ra}{\rightarrow}
\newcommand{\lra}{\longrightarrow}
\newcommand{\Lra}{\Leftrightarrow}
\newcommand{\bp}{\tilde \beta^\prime}
\newcommand{\Tr}{{\rm Tr} } 
\def\rme{{\rm e}}


\newfont{\namefont}{cmr10}
\newfont{\addfont}{cmti7 scaled 1440}
\newfont{\boldmathfont}{cmbx10}
\newfont{\headfontb}{cmbx10 scaled 1728}





\newcommand{\hyph}[1]{$#1$\nobreakdash-\hspace{0pt}}
\providecommand{\abs}[1]{\lvert#1\rvert}
\newcommand{\Nugual}[1]{$\mathcal{N}= #1 $}
\newcommand{\sub}[2]{#1_\text{#2}}
\newcommand{\partfrac}[2]{\frac{\partial #1}{\partial #2}}
\newcommand{\bsp}[1]{\begin{equation} \begin{split} #1 \end{split} \end{equation}}
\newcommand{\calF}{\mathcal{F}}
\newcommand{\calO}{\mathcal{O}}
\newcommand{\calM}{\mathcal{M}}
\newcommand{\calV}{\mathcal{V}}
\newcommand{\bbZ}{\mathbb{Z}}
\newcommand{\bbC}{\mathbb{C}}
\newcommand{\cK}{{\cal K}}

\newcommand{\Thq}{\Theta\left(\r-\r_q\right)}
\newcommand{\Dq}{\d\left(\r-\r_q\right)}
\newcommand{\kten}{\kappa^2_{\left(10\right)}}
\newcommand{\pbi}[1]{\imath^*\left(#1\right)}
\newcommand{\tth}{\tilde{\th}}
\newcommand{\tf}{\tilde{\f}}
\newcommand{\tj}{\tilde{\j}}
\newcommand{\tw}{\tilde{\omega}}
\newcommand{\tz}{\tilde{z}}
\newcommand{\prj}[2]{(\partial_r{#1})(\partial_{\j}{#2})-(\partial_r{#2})(\partial_{\j}{#1})}
\def\atanh{{\rm arctanh}}
\def\sech{{\rm sech}}
\def\csch{{\rm csch}}
\allowdisplaybreaks[1]

\def\red{\textcolor[rgb]{0.98,0.00,0.00}}

\newcommand{\Dan}[1] {{\textcolor{blue}{#1}}}

\numberwithin{equation}{section}



%

%
\setcounter{footnote}{0}
\renewcommand{\theequation}{{\rm\thesection.\arabic{equation}}}

\begin{titlepage}

\begin{center}

\vskip .5in 
\noindent

{\Large \bf{ From conformal to confining field theories using holography} }
\bigskip\medskip

Ali Fatemiabhari \footnote{a.fatemiabhari.2127756@swansea.ac.uk} and Carlos Nunez\footnote{c.nunez@swansea.ac.uk}\\

\bigskip\medskip
{\small 
Department of Physics, Swansea University, Swansea SA2 8PP, United Kingdom}

\vskip .5cm 
\vskip .9cm 
     	{\bf Abstract }\vskip .1in
\end{center}

\noindent
We construct a new family of Type IIB backgrounds that are dual to five dimensional conformal field theories compactified and deformed by VEVs of certain operators. This generates an RG flow into a smooth background dual to  non-SUSY gapped field theories in four dimensions. We study various holographic observables: a monotonic quantity associated with the number of degrees of freedom, Wilson loops that interpolate between conformal and confining behaviour with the possibility of screening, Entanglement Entropy, etc. We also give a prescription to compute the Holographic Complexity
in this type of backgrounds and calculate the spectrum of spin-two glueballs of the field theories. 
 \noindent
\vskip .5cm
\vskip .5cm
\vfill
\eject

\end{titlepage}

\setcounter{footnote}{0}

\small{
\tableofcontents}

\normalsize

\newpage
\renewcommand{\theequation}{{\rm\thesection.\arabic{equation}}}
\section{Introduction}
After the formulation of Maldacena's conjecture \cite{Maldacena:1997re} and the refinements in \cite{Gubser:1998bc},\cite{Witten:1998qj}, it became a natural project to extend the idea to Quantum Field Theories [QFTs] that are phenomenologically more appealing than ${\cal N}=4$ Super Yang-Mills. For example, Conformal Field Theories [CFTs] and QFTs with minimal or no-SUSY, to study the geometric realisation of phenomena like confinement, symmetry breaking, presence of condensates, etc.

Various works lead the way on this project: precursors were \cite{Itzhaki:1998dd}, \cite{Witten:1998zw}, \cite{Boonstra:1998mp}, followed by \cite{Girardello:1999hj}, \cite{Polchinski:2000uf} and others. The line of research represented by the papers
\cite{Klebanov:1998hh}-\cite{Dymarsky:2005xt} (see \cite{Gubser:2004tf} for a summary),
gave a very satisfactory geometric understanding of various non-perturbative aspects of a two node quiver field theory with ${\cal N}=1$ SUSY.
A different line of work based on wrapped branes 
\cite{Witten:1998zw},\cite{Maldacena:2000yy}, \cite{Acharya:2000mu}-\cite{Nunez:2001pt}
gave a complementary view on the same non-perturbative aspects (see  \cite{Aharony:2002up} for pedagogical reviews of this line of work). Satisfactorily, strong-coupling effects (for example, confinement, symmetry breaking, etc) have the same geometric realisation in both lines of work.
The last two lines of research were joined beautifully in the works \cite{Maldacena:2009mw},\cite{Gaillard:2010qg}, connecting both approaches.

A natural improvement on the previous lines of work (involving only adjoint or bifundamental fields) was the inclusion of fields transforming in the fundamental representation of the gauge group.
This was achieved first in the probe approximation, which included ''flavour branes'' probing the geometry (not backreacting, a sort of 'quenched' approximation). See \cite{Karch:2002sh}-\cite{Erdmenger:2007cm} for some representative papers and a review. This was later improved by including the backreaction of these sources. 
The flavour branes were 'smeared' in order to solve BPS ordinary differential equations rather than BPS PDEs. See for example \cite{Casero:2006pt}-\cite{Nunez:2010sf}.

A feature afflicting these models is the following: whenever the high energy behaviour of the QFT is field theoretical and represented by a deformed 4d CFT,  the IR  part of the holographic dual background is singular. This is the case for the models in \cite{Girardello:1999hj}, \cite{Polchinski:2000uf} and the more modern installments \cite{Bobev:2018eer}-\cite{Armas:2022bkh} (finite temperature alleviates the IR issue). On the other hand, the models with a smooth IR-part of the geometry have a UV regime that is typically not field theoretical, in the sense that the space is not asymptotically AdS.

Let us now focus on the contents of this paper, that attempts to partially remedy some of these unwelcomed features.
\subsection{General idea of this paper}
As anticipated, we attempt to address the undesirable feature mentioned above.
With this in mind, we consider a five dimensional quiver field theory. The QFT preserves eight Poincare supersymmetries and is balanced. The high energy dynamics leads to a strongly coupled conformal field theory. Holographically, the description of the five-dimensional fixed point is given in \cite{DHoker:2016ujz}-\cite{Legramandi:2021aqv}.

We then compactify the family of 5d SCFTs on a circle, giving anisotropic VEVs to $T_{\mu\nu}$ and a global symmetry current. An RG-flow ensues that ends in a four dimensional QFT with no-SUSY. The holographic dual for this flow across dimensions is in the framework (similar, but not identical) of solutions recently studied by Anabal\'on and Ross \cite{Anabalon:2021tua}. A precursor to these backgrounds can be found in \cite{Bobev:2020pjk} and further elaborations and applications found in the papers \cite{Anabalon:2022aig}-\cite{Nunez:2023xgl}.
We study holographically various aspects of the family of QFTs. The contents of this paper are distributed as follows. 

In Section \ref{sec:background}, we present a new family of Type IIB supergravity backgrounds, which serves as the basis for our study. The main characteristic is that it is smooth everywhere, except at the position of localised D7 brane sources (flavour branes). We calculate Page charges, finding that we are working with an array of D5-NS5 and D7 branes (as in the typical Hanany-Witten \cite{Hanany:1996ie} set up for five dimensional field theories). The number of branes present translates, as we discuss below, into the dual linear quiver being balanced.

In Section \ref{sec:QFT} we present a proposal for the dual QFT. As anticipated, the far UV is described by a five dimensional family of balanced linear quiver SCFTs. This is compactified on a  circle and deformed by an anisotropic VEV for $T_{\mu\nu}$ and for a global symmetry current, breaking SUSY. We study aspects of the low energy QFT using holography. Among these: a monotonic quantity associated with the number of degrees of freedom (we refer to it as  flow-central charge). This quantity indicates the presence of a 4d IR gapped system that is UV completed by a 5d SCFT. 

We also study the behaviour of Wilson loops, that start behaving as in a CFT for small separation of the non-dynamical quarks, and then this gives place to a  confining behaviour. By finding a suitable QCD string configuration, we make the point that the presence of localised sources (D7 flavour branes) gives the possibility of screening. The Entanglement Entropy on a rectangular strip is computed. We relate this quantity with the free energy of the 5d SCFT, and we study the contribution of the RG-flow to it.

We also propose a way to calculate
the Holographic Complexity for these types of flows across dimensions, away from SCFTs (in the C-V proposal \cite{Stanford:2014jda}). We find the complexity to be related to the free energy of the 5d SCFT and have contributions from the flow away from the fixed point. 

Finally, we compute the masses of spin-two excitations on the four dimensional Minkowski part of the space. The spectrum is of positive masses, giving some evidence that our non-SUSY QFT is actually stable.

We conclude in Section \ref{conclx}. We also give some possible ideas for further study that should be nice to understand. 
We have written detailed appendices explaining the calculations and various technical aspects. We hope these might be useful to colleagues wishing to work on these themes.

\section{The supergravity background}\label{sec:background}
In this section, we describe the new family of supergravity backgrounds studied in this paper and the associated brane charges. We refer the reader to Appendix \ref{appendix1} for detailed derivations.

The background is parameterised in terms of the coordinates  $\left( t,x_1,x_2,x_3,r, \phi,\theta,\varphi,\sigma,\eta \right) $, parameters $(c, \tilde{g},\mu)$, functions $H(r), f(r)$, seven functions $f_i(r,\sigma,\eta)$ and a function $V(\sigma,\eta)$. Setting  $\alpha'=g_s=1$, we have
\begin{align}
 d s^2_{st}& = f_1 \left[ \frac{2\tilde{g}^2}{9} H^{1/2}(r)\, r^2\,d \vec x _{1,3}+ \frac{2\tilde{g}^2}{9} \frac{H^{1/2}(r)}{f(r)}\, dr^2+ \frac{2\tilde{g}^2}{9} H(r)^{-3/2}\, f(r)\, d\phi^2  \right.\nonumber\\[2mm] 
& \left.+ f_2 \left(d \theta^2 +\sin{\theta}^2(d \varphi-A_1^{(3)})^2\right)+ f_3 \left(d \sigma^2+ d \eta^2 \right)\right]\, \label{tendconfig}\\[2mm]
f(r) &= -\fft{\mu}{r^3} + \ft29 \tilde{g}^2\, r^2\, H(r)^2\,,\qquad H(r)=1 - \fft{c^2}{\, r^3}\,,\quad A_1^{(3)} =  A_1^{(3)}(r) d\phi= \frac{\sqrt{2\mu}}{c}\,(1-\fr{1}{H(r)})
  \, d\phi\,,\nn\\[2mm]
 F_2^{(3)}& =dA^{(3)}_1= \frac{\sqrt{2\mu}}{c}\,\fr{H'(r)}{H(r)^2}dr\wedge d\phi,\;\;\;\; C_0 = f_7,\;\;\;\;  e^{-2\Phi} = f_6, \qquad F_5 = 4 (G_5 +*_{10} G_5), \nonumber\\[2mm]
 B_2& = f_4\text{Vol}(\tilde{S}^2) + \frac{2}{9} \eta \cos{\theta} F_2^{(3)} \,, \quad C_2 = f_5\text{Vol}(\tilde{S}^2)+4  \partial_\sigma (\sigma V) \cos{\theta} F_2^{(3)} \, . \nonumber\\
  \text{Vol}&(\tilde{S}^2)= \sin\theta d\theta \wedge \Big( d\varphi+ A_1^{(3)}(r) d\phi\Big).\nonumber
\end{align}
The physical significance of the parameters is described below. The functions $f_i(r,\sigma,\eta)$ are:
\begin{eqnarray}
& & f_1 =\frac{3\pi}{2 X^2} \left( \sigma^2 + \frac{3 X^4\sigma \partial_{\sigma}V}{\partial_{\eta}^2 V}\right)^{1/2}, \qquad f_2 = \frac{X^2\partial_{\sigma}V\partial_{\eta}^2V}{3{\Lambda}}, \qquad  f_3 = \frac{X^2\partial_{\eta}^2V}{3\sigma\partial_{\sigma}V} \label{defi} \\
& &   f_6= 12 \frac{{3 X^4 (\sigma^2 \partial_\sigma V) (\partial_{\eta}^2 V) }}{\left(3 X^4 \partial_\sigma V+\sigma  \partial_{\eta}^2 V\right)^2} \Lambda , \qquad f_7= 2 \left(\partial_\eta V+\frac{3 X^4 \sigma  \partial_\sigma V \partial_{\sigma \eta}^2V}{3 X^4 \partial_\sigma V+\sigma  \partial_{\eta}^2 V}\right), \nonumber\\
& & f_4= \frac{\pi}{2} \left(\eta -\frac{\sigma  \partial_{\sigma} V \partial_{\sigma \eta}^2 V}{\Lambda }\right),\qquad f_5=\frac{\pi}{2} \left(V-\frac{\sigma   \partial_{\sigma} V \left( \partial_{\eta} V \partial_{\sigma \eta}^2 V-3 X^4 \partial_{\eta}^2 V \partial_{\sigma} V \right)}{\Lambda}\right),\nonumber\\
& & \Lambda=3 X^4 \partial _{\eta}^2 V \partial _{\sigma }V+ \sigma \left[\left(\partial^2_{\eta \sigma} V \right)^2+\left(\partial _{\eta}^2 V\right)^2\right], \quad X(r) = \frac{1}{H(r)^{1/4}},~~~V=V(\sigma,\eta) ,\nonumber
\end{eqnarray}
$G_5$ is a differential form defined as

\begin{equation}
G_5= -\frac{4 \tilde{g}^2 \sqrt{2\mu} }{27 c} r^4 H'(r) f_1(r,\sigma ,\eta )  ~d t\wedge d x_1\wedge d x_2\wedge d x_3\wedge d \left(\cos\theta \sigma ^2 \partial_\sigma V\right) \, .
\end{equation}

The sphere $\tilde{S}^2$ is fibered over the $\phi$-coordinate. The 2d metric in  the 
$(\theta,\varphi)$ subspace reads
\begin{eqnarray}
& & ds^2_{\tilde{S}^2} = 
d \theta^2 +\sin^2{\theta}\Big(d \varphi-A_1^{(3)} \Big)^2
\end{eqnarray}

The circle parametrised by the angle $\phi$  shrinks smoothly at $r=r^*$ if we choose its periodicity to be 
\begin{eqnarray}\label{Period}
& & \phi\sim \phi +L_\phi,
~~    L_\phi\! =\!  \frac{8\pi \left(9 r^{*5}\right)^2 \left(1-\frac{c^2}{r^{*3}}\right)^{2}}{\left(-8 \tilde{g}^2 c^4 +4 \tilde {g}^2r^{*6}+4 \tilde {g}^2 c^2 r^{*3}+27 r^* \mu  \right)^2} \equiv 2\pi\left(\frac{2 H(r^*)}{f'(r^*)}\right)^2 .        \label{compactperiod}
\end{eqnarray}
The Ricci scalar is bounded for all the range of the radial coordinate $[r^*,\infty)$.

In the limit $r \rightarrow \infty$, the function  $H(r)\sim 1$, hence $
F^{(3)}_2(r\rightarrow\infty)\sim 0$. We impose that $
A^{(3)}_1(r)=0$ at the end of space $r=r^*$ by performing a large gauge transformation 
\begin{equation}
A_1^{(3)} =  \frac{\sqrt{2\mu}}{c}\,\left(1-\fr{1}{H(r)}-1+\fr{1}{H(r^*)}\right)
  \, d\phi\,= \frac{\sqrt{2\mu}}{c}\,\left(-\fr{1}{H(r)}+\fr{1}{H(r^*)}\right)
  \, d\phi \label{gaugefield}
\end{equation}
Of course,  $F_2^{(3)}$ remains unaltered. Note that as $r\to\infty$ the space approaches AdS$_6\times \tilde{S}^2(\theta,\varphi)\times\Sigma_2(\sigma,\eta)$.

After this gauge transformation, one can calculate the magnetic flux, 
that is related to the non-trivial holonomy of the gauge field
\begin{equation}
\Phi= - \oint A_\phi^{(3)} \left(  r=\infty\right)  d\phi= - \frac{1}{2}\int
{F^{(3)}}_{\mu\nu}dx^{\mu}\wedge dx^{\nu} =\frac{\sqrt{2\mu}}{c}\,\left(\fr{1}{H(r^*)}-1\right) L_{\phi}.
\end{equation}
Hence, the solution is described by two parameters; $\mu,c$ or $L_\phi, \Phi$, the latter being more natural parametrization from the boundary perspective\footnote{Note that there may be more than one background solution for a fixed periodicity $L_\phi$ and a magnetic flux $\Phi$ choice. Therefore choosing a pair of values for the pair $L_\phi, \Phi$ in the boundary may not fix a unique $\mu,c$  value in the bulk, leading to different branches of the background solutions. Here we are investigating one of them.}.

The two real coordinates, $(\sigma, \eta)$ parametrise a two dimensional Riemann surface $\Sigma$. To satisfy the type IIB equations of motion, the `potential function' $V(\sigma,\eta)$ should solve a Laplace-like differential equation with appropriate boundary conditions. The differential equations reads,
\begin{equation}
\partial_\sigma \left(\sigma^2 \partial_\sigma V\right) +\sigma^2 \partial^2_\eta V=0 \, .\label{diffeq} 
\end{equation}
To properly set the Laplace problem, it is convenient to define
\begin{equation}
V(\sigma,\eta)= \frac{\hat{V}(\sigma,\eta)}{\sigma}.
\end{equation}
In terms of $\hat{V}(\sigma,\eta)$ and the initial value function, a density of charge named ${\cal R}(\eta)$, the Laplace problem reads,
\begin{eqnarray}
& & \partial^2_\sigma \hat{V}+ \partial^2_\eta \hat{V} =0,\nonumber\\
& & \hat{V}(\sigma\to\pm\infty,\eta)=0,\;\;\;\;\;\hat{V}(\sigma, \eta=0)= \hat{V}(\sigma, \eta=P)=0.\nonumber\\
& & \lim_{\epsilon\to 0}\left(\partial_\sigma \hat{V}(\sigma=+\epsilon,\eta)- \partial_\sigma \hat{V}(\sigma=-\epsilon,\eta)\right)= {\cal R}(\eta).\label{bc}
\end{eqnarray}

All fields and warping factors depend on the  potential $V(\sigma,\eta)=\frac{\hat{V}(\sigma,\eta)}{\sigma}$ with support on the two-dimensional internal space.  
It can be shown that with eq.(\ref{diffeq}), all the equations of motion of the background in eq.(\ref{tendconfig}) are satisfied. In Appendix \ref{appendix2}, we  check the supersymmetry variations and show that the background is not preserving SUSY. The stability of the solution should be studied and we leave this for the future.

The configuration of eq.(\ref{tendconfig}) describes an infinite family of  asymptotically ($r\to\infty$) AdS$_6$ backgrounds in Type IIB.
The family is labelled by the function $V(\sigma,\eta)$ solving eq.(\ref{diffeq}). At high energies (for $r \rightarrow \infty$), the dual field theory develops a five dimensional fixed point. 

The potential function with appropriate boundary conditions, a solution to eq.(\ref{bc}), can be written in a Fourier expansion, as explained in \cite{Legramandi:2021uds}
\begin{equation}
\label{eq:fourier_vhat} 
 V(\sigma,\eta)= \frac{\hat{V} (\sigma,\eta)}{\sigma}, \quad\hat{V}(\sigma,\eta)= \sum_{k=1}^\infty a_k \sin\left(\frac{k\pi}{P}\eta \right) {e^{-\frac{k\pi}{P}|\sigma|}},\;\;\;\;a_k= \frac{1}{\pi k }\int_{0}^P {\cal R}(\eta)\sin\left( \frac{k\pi}{P} \eta\right) ~d\eta.
\end{equation}
As we discuss below, the  quantisation of Page charges forces the 'density of charge' or `Rank function' to  be a convex polygonal,
\begin{equation}
     {\cal R}(\eta) = \begin{cases} 
          N_1 \eta & 0\leq \eta \leq 1 \\
          N_l+ (N_{l+1} - N_l)(\eta-l) & l \leq \eta\leq l+1,\;\;\; l:=1,...., P-2\\
  %
          N_{P-1}(P-\eta) & (P-1)\leq \eta\leq P . 
       \end{cases} \label{eq:Rank}
\end{equation}
Note that the variable $\eta$ is bounded in the interval $[0,P]$ and $\sigma$ ranges over the real axis $(-\infty,\infty)$. 

We now study the behaviour of this family of solutions close to special points in the geometry of the internal space.
%


\subsection{Behaviour at special points}

We investigate the behaviour of the metric and the dilaton as we approach special points. Specially we will focus on the points $\eta= 0, \eta= P$, $\sigma \to \pm \infty$, with $r \to \infty$.

We start by analysing the metric behaviour on the boundary at $\eta=0$ with  identical results for $\eta =P$. We set $r \to \infty$,  which implies  $X(r) \to 1$. In this boundary, $f_1$ and $f_3$ are finite. Explicitly,
\begin{align}
f_1^2(r\to\infty,\sigma,0) =&  \frac{9\pi^2}{4} \, \, \frac{\sum _{k=1}^\infty  \frac{\pi k}{P} a_k \left(\left(\frac{\pi  k \sigma }{P}\right)^2+3 | \sigma | \frac{\pi  k \sigma }{P} +3\right) e^{-\frac{\pi  k | \sigma | }{P}}}{\sum _{k=1}^\infty a_k \left(\frac{\pi  k}{P}\right)^3 e^{-\frac{\pi  k | \sigma | }{P}}} \, , \nonumber\\[2mm]
f_3(r\to\infty,\sigma,0) =& \frac13 \, \, \frac{ \sum _{k=1}^\infty a_k \left(\frac{\pi  k}{P}\right)^3 e^{-\frac{\pi  k | \sigma | }{P}}}{ \sum _{k=1}^\infty a_k \left(\left(\frac{\pi  k}{P}\right)^2 | \sigma | + \frac{\pi k}{P} \right) e^{-\frac{\pi  k | \sigma | }{P}}} \, , \nonumber
\end{align}
while $f_2 \to \eta^2 f_3(+\infty, \sigma,0)$. Using these results at these boundaries, the metric is
\begin{equation}
\begin{split}
\eta \to 0 \, \qquad d s^2_{10} =& f_1(+\infty, \sigma,0) \, \Big(d s^2 (\text{AdS}_6) + f_3(+\infty, \sigma,0) (\eta^2 d s^2 (\tilde{S}^2)+  d \eta^2 + d \sigma^2 ) \Big) \, ,\\
\eta \to P \, \qquad d s^2_{10} =& f_1(+\infty, \sigma,P) \Big(d s^2 (\text{AdS}_6) + f_3(+\infty, \sigma,P) ((\eta-P)^2 d s^2(\tilde{S}^2) +  d \eta^2 + d \sigma^2  ) \Big) \, .
\end{split}
\end{equation}
In the limit $r \to \infty$, $A_1^{(3)} \to \frac{\sqrt{2\mu}}{c H(r^*)}d\phi$, hence the fibration of $\tilde S^2$ on AdS$_6$ is trivial.

The metric is regular in these two limits as it is given by a warped product AdS$_6 \times \mathbb{R}^4$ with generally non-singular warpings. The regularity of the dilaton, being finite, confirms the regularity of the solution at these boundaries. 

Now we consider the limit $\sigma \to \pm \infty$, $r \to \infty$. Using eq.(\ref{eq:fourier_vhat}), the leading contribution to  the potential $\hat{V}=\sigma V$ comes from the mode with $k=1$. The asymptotic expansions needed to study the space-time are,
\begin{equation}
\sigma V(\sigma,\eta) \sim  \partial^2_\eta (\sigma V) \sim \sin\left(\frac{\pi}{P}\eta \right)  e^{-\frac{\pi}{P}|\sigma|} \, , \quad \sigma^2 \partial_\sigma V \sim |\sigma| \sin\left(\frac{\pi}{P}\eta \right)  e^{-\frac{\pi}{P}|\sigma|} \, , \quad \Lambda \sim \sigma^{-1} e^{-\frac{2\pi}{P}|\sigma|} \, .
\end{equation}
Making use of these relations and up to constant factors, one has
\begin{equation}
\sigma \to \pm \infty \, \qquad d s^2 = |\sigma| d s^2 (\text{AdS}_6) + \sin^2 \left( \frac{\pi}{P} \eta\right) d s^2 (S^2)+  d \eta^2 + d \sigma^2 .
\end{equation}
With a similar analysis, the dilaton reads
\begin{equation}
\sigma \to \infty \qquad  e^{-\Phi} \sim \frac{ e^{- \frac{\pi}{P} |\sigma|}}{\sqrt{|\sigma|}} \, .
\end{equation}
By the change of coordinates $|\sigma| \to -\log z$ with $z$ positive and small, the metric and the dilaton take the form of a $(p,q)$-five-brane in $\sigma\to\pm\infty$, as described in \cite{DHoker:2016ujz}.

The behaviour of this family of 6d backgrounds at $\sigma = 0$ is characterised by the presence of singularities in the fluxes, dilaton and metric. We calculate the conserved Page charges to study these behaviours. As we discuss below, thanks to constraints we imposed on the Rank function
${\cal R}(\eta)$--that is that the density of charge is a convex polygonal as given in eq.(\ref{eq:Rank}), the quantisation conditions for charges are satisfied.

\subsection{Page charges}\label{pagesection}
%
%

In this section, we analyse the conserved and quantised Page charges in the background of eq.(\ref{tendconfig}). Below, we find that the restriction on the function ${\cal R}(\eta)$ in eq.(\ref{eq:Rank}) 
to be a convex, piece-wise linear function implies the quantisation of the charges.

In our conventions, the volume element of the sphere is $\text{Vol}(\tilde S^2)= \sin\theta d\theta\wedge d\varphi- \sin\theta A_1^{(3)}(r) d\theta\wedge d\phi$, hence the field strengths read
\begin{eqnarray}
& & H_3= dB_2= d\big[f_4\text{Vol}(\tilde{S}^2) + \frac{2}{9} \eta \cos{\theta} F_2^{(3)} \big] =\left( \partial_\sigma f_4 d\sigma +\partial_\eta f_4 d\eta \right)\wedge \text{Vol}(\tilde S^2) + \cdots,\; \label{fieldstrength}\\[2mm]
& & \hat{F}_1= F_1=dC_0= \partial_\sigma f_7 d\sigma+\partial_\eta f_7 d\eta+\partial_r f_7 d r, \nonumber\\[2mm]
& &  \hat{F}_3= F_3- B_2\wedge F_1= d\left( C_2 - C_0 B_2\right)= \left[ \partial_\sigma (f_5- f_7 f_4) +\partial_\eta(f_5-f_7 f_4)\right]\wedge \text{Vol}(\tilde S^2)+ \cdots. \nonumber
\end{eqnarray}
We have used the definition of the Page fluxes $\hat{F}= F \wedge e^{-B_2}$, and we have only kept terms in the expressions that are relevant for our calculation of the charges.

Setting $\alpha'=g_s=1$ we have,
\begin{equation}
 Q_{Dp,Page}=\frac{1}{(2\pi)^{7-p}}\int_{\Sigma_{8-p}}\hat{F}_{8-p}.\nonumber
\end{equation}
This implies,
\begin{eqnarray}
& & Q_{NS5}=\frac{1}{4\pi^2}\int_{M_3} H_3,\;\;\;\; Q_{D7}= \int_{\Sigma_1} \hat{F}_1,\;\;\; Q_{D5}=\frac{1}{4\pi^2 } \int_{\Sigma_3} \hat{F}_3.\label{pagecharges}
\end{eqnarray}
The cycles $M_3, \Sigma_1, \Sigma_3$ are defined as,
\begin{eqnarray}
& & M_3=[\eta, S^2], \textrm{with}~\sigma\to\pm\infty,~r\to\infty,\;\;\;\;\; ~\Sigma_1=[\eta], \textrm{with}~\sigma=0~, r\to\infty, 
\nonumber\\
& & \Sigma_3=[\sigma, S^2], \textrm{with}~\eta=\textrm{fixed},~r\to\infty.\nonumber
\end{eqnarray}
We allow the possibility of a large gauge transformation $B_2\to B_2 + \Delta d\Omega_2$. This does not alter the charge of NS5 or the D7 brane charge, but it has an effect on the 
charge of D5 branes. 

On the other hand, the field strength $\hat{F}_5=F_5- B_2\wedge F_3+\fr1 2 B_2\wedge B_2 \wedge F_1$ does not induce quantised D3 brane charges as 
all possible five-cycles are non-compact.
Let us analyse the three possible quantised charges.
\\

\underline{\bf NS-5 branes:}
Calculating explicitly the NS five branes charge, we have
\begin{equation}
\pi Q_{NS5}=\frac{1}{4\pi}\int_{M_3} H_3= \int d\eta \partial_{\eta} f_4(r\to\infty, \sigma\to\pm\infty,\eta)= f_4(\infty, \pm\infty,P)- f_4(\infty, \pm\infty,0) .
\end{equation} 
Using the expressions in eqs.(\ref{B2identity})-(\ref{identityB2}),
we find that the number of NS-five branes is
\begin{equation}
Q_{NS5}= 
P.\label{QNS5}
\end{equation}
Note that both the contribution of the NS-five branes coming from $\sigma=+\infty$ and $\sigma=-\infty$ are included. This indicates that $P$ NS-five branes exist in the background. 

We could consider another cycle 
\begin{eqnarray}
& & M'_3=[\eta, \theta, \phi], \textrm{with}~r\to \infty, ~\sigma\to \pm\infty,\;\;\;.\nonumber
\end{eqnarray}
with the NS five brane charge,
\begin{align}
\pi Q'_{NS5}=\frac{1}{4\pi}\int_{M'_3} H_3&= -\int d\eta A_1^{(3)}(r=\infty)\partial_{\eta} f_4(r\to\infty,\sigma\to\pm\infty,\eta) \nonumber \\
&=-A_1^{(3)}(r=\infty)\left(f_4(\infty,\pm\infty,P)- f_4(\infty, \pm\infty,0)\right),\nonumber \\
Q'_{NS5}&= -A_1^{(3)}(r=\infty)P.\label{QNS52}    
\end{align}
To have this charge quantised, we must quantise the $c$-parameter appearing in eq.(\ref{tendconfig})---see eq.(\ref{gaugefield}). Notice that the cycle $M'_3$ is not topologically $S^3$ but $S^2\times S^1$.
Now we check the D7 brane charge.
\\
\\
\underline{\bf D7 branes:}
For the charge of D7 branes, we have,
\begin{eqnarray}
Q_{D7}= \int_{\Sigma_1} \hat{F}_1= \int_{0}^P d\eta \partial_\eta f_7(\infty, 0,\eta)= f_7(\infty, 0,P)- f_7(\infty, 0,0).
\end{eqnarray}
By using  the identity in eq.(\ref{identityC0}), we have
\begin{equation}
Q_{D7}= \left( {\cal R}'(0) -{\cal R}'(P)\right).
\end{equation}
In the first interval $0\leq\eta\leq 1$ we have ${\cal R}= N_1\eta$ and in the last interval
$P-1\leq \eta\leq P$ we have ${\cal R}= N_{P-1} (P-\eta)$, as a result of integer slopes. The number of D7 branes reads
\begin{equation}
Q_{D7}=  (N_1+ N_{P-1}).\label{numberofd7}
\end{equation} 

\underline{\bf D5 branes:}
We perform a large gauge transformation on $B_2\to B_2 + \Delta \text{Vol}(S^2)$. Here $\text{Vol}(S^2)= \sin\theta d\theta\wedge d\varphi$ which is well-defined volume element of a sphere $S^2$ in $r \to \infty$ of our metric. The charge of D5 branes is
\begin{eqnarray}
& & \pi Q_{D5}=\frac{1}{4\pi}\int_{\Sigma_3}F_3- (B_2 +\Delta  \text{Vol}(S^2))\wedge F_1= \int_{-\infty}^{\infty} d\sigma \partial_\sigma \left[f_5- f_7(f_4+\Delta) \right]=\nonumber\\
& & \int_{-\infty}^{-\epsilon} \partial_\sigma [f_5- f_7(f_4+\Delta) ]+\int_{\epsilon}^\infty \partial_\sigma [f_5- f_7(f_4+\Delta) ].
\end{eqnarray}
The quantity $f_5- f_7(f_4 +\Delta)$ needs to be evaluated at $\sigma\to\pm\infty$ and $\sigma=\pm \epsilon$ then take $\epsilon\to 0$. 
The details of calculations are given in eqs.(\ref{c2b2c01})-(\ref{c2b2c02}). The combination $f_5- f_7 (f_4 +\Delta)$ vanishes at $\sigma\to\pm\infty$. 
Hence we only need to calculate
\begin{equation}
\pi Q_{D5}= f_5-f_7(f_4+\Delta) \Big]_{\epsilon}^{-\epsilon}
\end{equation}
evaluated at some fixed value of the $\eta$-coordinate and $r \to \infty$.
Using the expressions in eqs.(\ref{identityC0}) and (\ref{c2b2c02}) we find
\begin{equation}
Q_{D5}= {\cal R}(\eta) -{\cal R}'(\eta) \left(\eta-\Delta\right).\label{Qd5}
\end{equation}
Remind that in each interval, the function ${\cal R}(\eta)$ in eq.(\ref{eq:Rank})  is linear, with integer intercept and slope, hence $Q_{D5}$ is an integer. In fact, choosing an interval-dependent large gauge transformation $\Delta=k$, in the $[k,k+1]$ interval
 eq.(\ref{Qd5}) gives 
\begin{equation}
Q_{D5}= N_k.\label{Qd5final}
\end{equation}
We interpret this by relating the node in the quiver with gauge group $SU(N_k)$ with the interval $[k, k+1]$ of the $\eta$-coordinate.

Similar to the discussion in the NS5 charge calculation, one can consider another cycle 
\begin{eqnarray}
& & \Sigma'_3=[\sigma, \theta, \phi], \textrm{with}~\eta=\textrm{fixed},~r\to\infty\;\;\;.\nonumber
\end{eqnarray}
with the D5 branes charge,
\begin{eqnarray}
& & \pi Q'_{D5}=\frac{1}{4\pi}\int_{\Sigma'_3}F_3- (B_2 +\Delta  \text{Vol}(S^2))\wedge F_1= -\int_{-\infty}^{\infty} d\sigma A_1^{(3)}(r=\infty)\partial_\sigma \left[f_5- f_7(f_4+\Delta) \right]=\nonumber\\
& & -A_1^{(3)}(r=\infty)\left(f_5-f_7(f_4+\Delta) \right)\Big]_{\epsilon}^{-\epsilon}  \nonumber \\
& & Q'_{D5}= -A_1^{(3)}(r=\infty) N_k
\end{eqnarray}
As above, in order to have this charge quantised charge, one must impose a quantization condition on the $c$-parameter in eq.(\ref{tendconfig})---see also eq.(\ref{gaugefield}). The cycle $\Sigma'_3$ is not topologically $S^3$ but $S^2\times S^1$.

To summarise, in this background, the total number of the branes in the system is given by
\begin{eqnarray}
& & Q_{NS5} = P \, \label{chargesfinal}\\
& &  Q_{D7}[k, k+1]= {\cal R}''(k)=(2 N_{k} - N_{k+1}- N_{k-1}),\; Q_{D7,total}=(N_1+ N_{P-1})= \int_0^P {\cal R}'' (\eta) d \eta ,\nonumber\\
 & & Q_{D5}[k,k+1] = {\cal R}(\eta) -{\cal R}'(\eta) (\eta- \Delta) =N_k\, ,\;\;\;\;\; Q_{D5,total}=\int_0^P {\cal R} ~d\eta.\nonumber 
\end{eqnarray}

We will now discuss the associated QFTs. With the presence of branes in our configuration, specifically, D7 sources, for our background to be trustable, we must consider large values of $P$. This  ensures that the D7 flavour branes (generating singularities in the background) are separated enough. In other words, the backgrounds in eq.(\ref{tendconfig}) are dual to field theories that in the UV consist of long linear quivers with sparse flavour groups.

\section{Dual Field Theories and Observables}\label{sec:QFT}

In this section, we discuss the field theories dual to the family of type IIB backgrounds presented in Section \ref{sec:background}. 

The procedure to determine the holographic dual unfolds as follows.
For large values of the radial coordinate, $r \rightarrow \infty$, the backgrounds asymptote to AdS$_6$ as the field $X(r\rightarrow\infty) \sim 1$,  the field $
A^{(3)}_1(r\rightarrow\infty)\approx  \frac{2\sqrt{\mu}}{c H(r^*)} d\phi$, that is a pure gauge field-see eq.(\ref{gaugefield}), and $
F^{(3)}_2(r\rightarrow\infty)\sim 0$.
Due to the presence of fluxes and a non-trivial $S^2$ fibration over the $\phi$-coordinate, the AdS$_6$ isometries are altered as we move inside the bulk, towards $r^*$. This has some resemblance with the twisted compactifications described in \cite{Legramandi:2021aqv} (see also \cite{Merrikin:2022yho}). In the cases studied in \cite{Legramandi:2021aqv}, an infinite family of 5d SCFTs, characterized by functions $V(\sigma, \eta)$ that satisfy a Laplace equation, are compactified on a curved manifold. Twisted compactifications were initially introduced by Witten and comprehensively reviewed in \cite{Witten:1994ev}. For studies in different contexts and details across numerous examples, the reader is referred to \cite{Maldacena:2000mw, Acharya:2000mu, Nunez:2001pt, Bobev:2017uzs, Kim:2019fsg}. In the present case, the compactification on $S^1_\phi$ does not lead to the preservation of SUSY, as massless spinors  are not  supported in our background (See Appendix \ref{appendix2} for more details). In this sense, the solution in eq.(\ref{tendconfig}) is not a twisted compactification. Also, the manifold on which we compactify, $S^1_\phi$, is not curved.

In the present context, we see our backgrounds as providing a holographic representation of the compactification of five-dimensional QFTs on a circle with a Wilson line added-represented holographically by the $A^{(3)}_1$-fibre in eq.(\ref{gaugefield}). 
The type of backgrounds we work with belong to a class of recently studied solutions by Anabal\'on and Ross in \cite{Anabalon:2021tua}. A precursor to these backgrounds can be found in \cite{Bobev:2020pjk} and further elaborations and applications found in the papers \cite{Anabalon:2022aig}-\cite{Nunez:2023xgl}.

These five-dimensional linear quivers approach a conformal fixed point at high energies. The UV-SCFT is deformed by introducing operators that describe the compactification.

In addition, a QFT Wilson line, holographically implemented by the field $A^{(3)}_1$ in eq.(\ref{gaugefield}), is switched on in the QFT. The insight provided by our geometries in eqs.(\ref{tendconfig})-(\ref{defi}) is that at lower energies relative to the  finite size of the compact space $S_\phi^1$, the field theories undergo a transition to non-conformal field theories in $(3+1)$ dimensions.  

The five dimensional CFTs that we compactify describe the strongly coupled dynamics (at high energies) of  linear (balanced) quiver field theories, as those in Figure \ref{fig:quiver}.
\begin{figure}
\begin{center}
	\begin{tikzpicture}
	\node (1) at (-4,0) [circle,draw,thick,minimum size=1.4cm] {N$_1$};
	\node (2) at (-2,0) [circle,draw,thick,minimum size=1.4cm] {N$_2$};
	\node (3) at (0,0)  {$\dots$};
	\node (5) at (4,0) [circle,draw,thick,minimum size=1.4cm] {N$_{P}$};
	\node (4) at (2,0) [circle,draw,thick,minimum size=1.4cm] {N$_{P-1}$};
	\draw[thick] (1) -- (2) -- (3) -- (4) -- (5);
	\node (1b) at (-4,-2) [rectangle,draw,thick,minimum size=1.2cm] {F$_1$};
	\node (2b) at (-2,-2) [rectangle,draw,thick,minimum size=1.2cm] {F$_2$};
	\node (3b) at (0,0)  {$\dots$};
	\node (5b) at (4,-2) [rectangle,draw,thick,minimum size=1.2cm] {F$_P$};
	\node (4b) at (2,-2) [rectangle,draw,thick,minimum size=1.2cm] {F$_{P-1}$};
	\draw[thick] (1) -- (1b);
	\draw[thick] (2) -- (2b);
	\draw[thick] (4) -- (4b);
	\draw[thick] (5) -- (5b);
	\end{tikzpicture}
\end{center}
    \caption{A linear quiver. The balancing condition implies $F_i = 2 N_i - N_{i-1}-N_{i+1}$}
    \label{fig:quiver}
\end{figure}
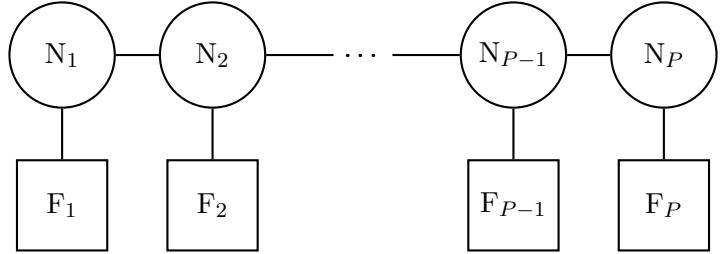
The function $V(\sigma,\eta)$ is uniquely determined by the numbers $N_1, N_2,....,N_P$ and $F_1, ...., F_P$. 
In fact, these numbers determine the function ${\cal R}(\eta)$, which using eq.(\ref{eq:fourier_vhat}) determines $V(\sigma,\eta)$.

These UV conformal points  are deformed by relevant operators. One can  read the dimension of these operators from the near-AdS$_6$ expansion of the gauged supergravity metric--see Appendix \ref{appendix1} for the details of such six dimensional system and Appendix \ref{appendix:Boundary} for details of the near boundary analysis.

The parameters $\mu, c$ control the subleading modes of the metric, scalar field $X$ and the  gauge field in the asymptotic expansions. Hence, they are interpreted as VEVs of the corresponding dual operators on the field theory side. The asymptotic expansions are presented in detail in the Appendix \ref{appendix:Boundary}. The gauge field $A_1^{(3)}$ introduces a VEV for a global symmetry current in the dual theory. The field $X$ leads to the VEV of an operator of dimension three. The VEVs are,
\begin{equation}
    \langle J \rangle = -3c\sqrt{\mu}  , \quad  \langle {\cal O}_X \rangle = \frac{c^2}{4},\label{VEVs1}
\end{equation}
respectively (we have set $\gt=\sqrt{\frac{9}{2}}$ in this section). The field $A_1^{(3)}$ couples to a background global symmetry current in the boundary. This can be interpreted as a background Wilson loop insertion in the QFT.
The VEV for the components of the boundary stress tensor are,
\begin{equation} 
    \langle T_{tt} \rangle = -\mu , \quad  \langle T_{x_i x_i} \rangle = \mu , \quad \langle T_{\phi\phi} \rangle = 4\mu.\label{VEVs2}
\end{equation}
Note that the energy density of the solution is negative in the case of positive $\mu$. The VEVs in the compact $\phi$-direction and the other flat $x_i$-directions are different, hence the insertion is anisotropic. 
These relevant operators trigger an RG flow, which ends in a set of gapped 4d QFTs dual to the backgrounds at hand. In Figure \ref{fig:RG}, we depict the field theory structure described here under the RG flow.

\begin{figure}
\begin{flushright}
\begin{tiny}
	\begin{tikzpicture}
 \def \zzz {0.8}
	\node (1) at (-4*\zzz ,0*\zzz ) [circle,draw,thick,minimum size=1.4*\zzz cm] {N$_1$};
	\node (2) at (-2*\zzz ,0) [circle,draw,thick,minimum size=1.4*\zzz cm] {N$_2$};
	\node (3) at (0,0)  {$\dots$};
	\node (5) at (4*\zzz ,0) [circle,draw,thick,minimum size=1.4*\zzz cm] {N$_{P}$};
	\node (4) at (2*\zzz ,0) [circle,draw,thick,minimum size=1.4*\zzz cm] {N$_{P-1}$};
 \node (5d) at (6*\zzz ,-1*\zzz )  {\small \bf 5d QFT};
	\draw[thick] (1) -- (2) -- (3) -- (4) -- (5);
	\node (1b) at (-4*\zzz ,-2*\zzz ) [rectangle,draw,thick,minimum size=1.2*\zzz cm] {F$_1$};
	\node (2b) at (-2*\zzz ,-2*\zzz ) [rectangle,draw,thick,minimum size=1.2*\zzz cm] {F$_2$};
	\node (3b) at (0,0)  {$\dots$};
	\node (5b) at (4*\zzz ,-2*\zzz ) [rectangle,draw,thick,minimum size=1.2*\zzz cm] {F$_P$};
	\node (4b) at (2*\zzz ,-2*\zzz ) [rectangle,draw,thick,minimum size=1.2*\zzz cm] {F$_{P-1}$};
	\draw[thick] (1) -- (1b);
	\draw[thick] (2) -- (2b);
	\draw[thick] (4) -- (4b);
	\draw[thick] (5) -- (5b);
  \node (A) at (7.5*\zzz ,-1*\zzz ) {};
  \node (B) at (7.5*\zzz ,6*\zzz ) {};
   \node (En) at (7.5*\zzz ,7*\zzz )  {\small Energy};
   \node (C) at (0*\zzz ,1*\zzz ) {};
  \node (D) at (0*\zzz ,5*\zzz ) {};
  \node (5dS) at (0*\zzz ,6*\zzz )  [rectangle,draw,thick,minimum size=1.2*\zzz cm] {\small \bf 5d SCFT};
   \node (5dSWe) at (-1.5*\zzz ,6*\zzz ) {};
  \node (E) at (-10*\zzz ,1.5*\zzz ) [rectangle,draw,thick,minimum size=1.2*\zzz cm] {\small \bf 4d QFT};
  \node (EEa) at (-8.6*\zzz ,2.4*\zzz ) {};
  \node (T1) at (-5*\zzz ,6*\zzz ) {\small Deformations};
  \node (T2) at (-5*\zzz ,5*\zzz ) {\small $\langle T \rangle, \langle J \rangle, \cdots$ };
 \path[->,>=stealth',auto, node distance=3cm,
  thick, main node/.style={draw,font=\sffamily\Large\bfseries} , every node/.style={font=\sffamily\small}]
    (A) edge node [left] {} (B)
    (C) edge node [left] {} (D)
    (5dSWe) edge node [left] {} (EEa);
	\end{tikzpicture}
 \end{tiny}
    
\end{flushright}
    \caption{Structure of the 5d SCFTs under RG flow}
    \label{fig:RG}
\end{figure}
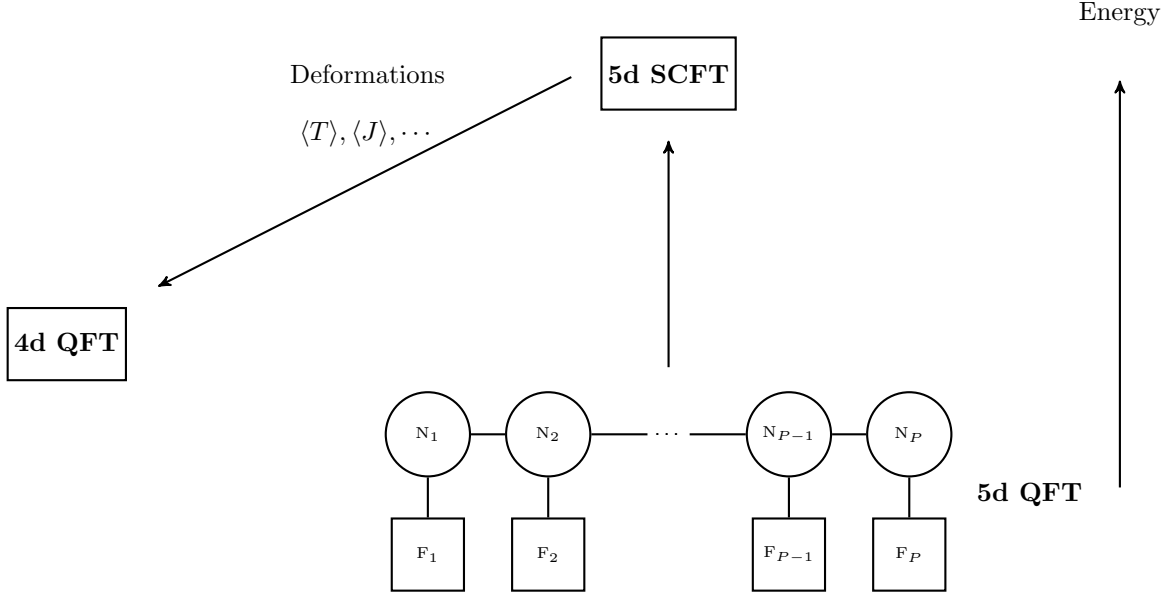

In the next sections we calculate various observable quantities of the QFT. Starting with a quantity called holographic central charge in Section \ref{sec:holcent}. The holographic central charge is defined to measure the number of degrees of freedom (or the Free Energy) of these strongly coupled lower dimensional QFTs. As we find below, the result is described in terms of a function of the energy scale and transcendental functions of the parameters in the quiver, revealing the non-perturbative character of the QFTs.

\subsection{The holographic central charge}\label{sec:holcent}

In this section, we calculate the holographic central charge, a quantity that gives a measure of the number of degrees of freedom (number of states) along the flow. 
Following \cite{Bea:2015fja}, \cite{Macpherson:2014eza}, we start defining the quantity  at conformal points and explicitly compute it in a couple of examples of our type IIB family of geometries.
This contains information about the free energy or the number of degrees of freedom in the strongly coupled 5d CFT.

After studying the conformal UV
we present the central charge along the flow, originally defined in  \cite{Bea:2015fja}. Being applicable to geometries describing flows, its main characteristic is that it is constant at both ends of the flow. 

\subsubsection{The holographic central charge at fixed points ($r\to\infty$)}\label{choldef}
We briefly summarise the holographic central charge formalism.
Consider a $(d+1)$ dimensional QFT dual to a background, with metric  and dilaton of the form,
\begin{eqnarray}
ds^2= \alpha(r,\vec{\theta})\Big(dx_{1,d}^2 + \beta(r) dr^2 \Big) + g_{ij}(r,\vec{\theta}) d \theta^i d \theta^j,\;\;\;\;\; ~~~~\Phi(r,\vec \theta), \label{assignation}
\end{eqnarray}
One calculates $V_{int}$, a weighted version of the internal volume defined below, and with this, the central charge,
\begin{eqnarray}
& & V_{int}=\int d \vec{\theta}\sqrt{\det[g_{ij}] e^{-4\Phi} \alpha^d},\;\;\;\; \hat{H}= V_{int}^2,\nonumber\\
& & c_{hol}= \frac{d^d}{G_N}\beta^{d/2}\frac{\hat{H}^{\frac{2 d+1}{2}}}{(\hat{H}')^d}\label{chol}
\end{eqnarray}
We follow the above procedure for the UV fixed point solution, the $AdS_6$ backgrounds arising as $r\to\infty$ which are dual to the far UV limit of our QFTs. We set  $d=4$ in eqs.(\ref{assignation}), (\ref{chol}) and choose  Poincar\'e coordinates for $AdS_6$. Comparing with eq.(\ref{tendconfig}) in the limit of large-$r$ and after obvious coordinates redefinitions, we find,
\begin{eqnarray}
& & ds^2(\text{AdS}_6) = r^2 dx_{1,4}^2+ \frac{dr^2}{r^2},\;\;\;\alpha= f_1(\sigma,\eta) r^2,\;\;\;\;\beta=\frac{1}{r^4},~ d=4 \\
& &g_{ij}d\theta^i d\theta^j= f_1 f_2 d\tilde{S}^2+ f_1 f_3 (d\sigma^2+d\eta^2),~~ V_{int }= {\cal N} r^4,~~~ {\cal N}=  \int d\theta~d\varphi~ d\sigma~ d \eta~~ \sin\theta~ f_1^4 f_2 f_3 f_6.\nonumber
\end{eqnarray}
Using eq.(\ref{chol}) and the functions $f_i(r\to\infty,\sigma,\eta)$ and $X(r\to\infty)=1$ in eq.(\ref{defi}), gives the holographic central charge,
\begin{eqnarray}
& & c_{hol}= \frac{1}{16 G_N}{\cal N}, \label{chol2} \\
& & 
G_N= 8\pi^6, ~~
{\cal N}= 3^3\pi^5 \int_{0}^{P} d \eta \int_{-\infty}^{\infty} d \sigma~\sigma^3 \partial_\sigma V \partial_\eta^2 V .\label{vavar}
\end{eqnarray}
\\
\underline{\bf Examples}
\\
Now, we present two case study examples. In each case we write the rank function ${\cal R}(\eta)$, the Fourier coefficient $a_k$ and the Potential $\hat{V}(\sigma,\eta)$ in eq.(\ref{eq:fourier_vhat}). We show the associated 5d quiver field theory and calculate the holographic central charge. 

Let us first consider a gauge theory called $\tilde{T}_{N,P}$. The gauge theory in the IR is described  
by the quiver 
\begin{center}
	\begin{tikzpicture}
	\node (1) at (-6,0) [circle,draw,thick,minimum size=1.4cm] {N};
	\node (2) at (-4,0) [circle,draw,thick,minimum size=1.4cm] {2N};
	\node (3) at (-2,0) [circle,draw,thick,minimum size=1.4cm] {3N};	
	\node (4) at (0,0)  {$\dots$};
	\node (6) at (4,0) [rectangle,draw,thick,minimum size=1.2cm] {PN};
	\node (5) at (2,0)  {(P-1)N};
	\draw[thick] (1) -- (2) -- (3) -- (4) -- (5)-- (6);
	\draw[thick] (2,0) circle (0.7cm) ;
	\draw[thick] (1,0) -- (1.3,0);
	\draw[thick] (2.7,0) -- (3.3,0);
	\end{tikzpicture}\
\end{center}
The rank function corresponding to this quiver is,
\[ {\cal R}(\eta) = \begin{cases} 
N\eta & 0\leq \eta \leq (P-1) \\
N(P-1) (P-\eta)& (P-1)\leq \eta\leq P .
\end{cases}
\]
The  coefficient $a_k$ and the potential $\hat{V}(\sigma,\eta)$ defined in eq.(\ref{eq:fourier_vhat}) are,
\begin{eqnarray}
& &  a_k= (-1)^{k+1}\frac{N P^3}{k^3 \pi^3} \sin\left( \frac{k\pi }{P}\right). \label{ak1}\\
& & \hat{V} = \frac{N P^3}{2 \pi ^3} \text{Re} \left(\text{Li}_3(-e^{-\frac{\pi}{P}  (| \sigma |+i+i \eta  )})-\text{Li}_3(-e^{-\frac{\pi}{P}  (| \sigma |-i+i \eta )}) \right) \, . \label{eq:V_TN}
\end{eqnarray}
Using eq.(\ref{chol2}), the holographic central charge for this theory reads 
\begin{equation}
c_{hol}= \frac{N^2 P^6}{8\pi^{10}}\left(2\zeta(5) -{\text{Li}_5(e^{\frac{2\pi i}{P}} ) -\text{Li}_5(e^{-\frac{2\pi i}{P}})} \right).
\end{equation}
This result is trustable for long quivers. Thus, we take the $P\to\infty$ limit and to the leading order we find,
\begin{equation}
c_{hol}=\frac{N^2 P^4}{2\pi^8}\zeta(3) +O\left(\frac{\log P}{P^2}\right).\label{example1}
\end{equation}
The presence of $\zeta(3)$ indicates the intrinsically non-perturbative character of this quantity and of the associated 5d SCFT. This result can be reproduced with a Matrix Model approach as shown in \cite{Uhlemann:2019ypp,Akhond:2022awd,Akhond:2022oaf,Nunez:2023loo}.

We investigate a second  example, known as the $+_{P,N}$ theory. The rank function is given by,
\[ {\cal R}(\eta) = \begin{cases} 
N\eta & 0\leq \eta \leq 1 \\
N & 1\leq \eta\leq (P-1)\\
N (P-\eta) & (P-1)\leq \eta\leq P .
\end{cases}
\]
This is equivalent to a linear quiver field theory,
\begin{center}
	\begin{tikzpicture}
	\node (1) at (-4,0) [rectangle,draw,thick,minimum size=1.2cm] {N};
	\node (2) at (-2,0) [circle,draw,thick,minimum size=1.4cm] {N};
	\node (3) at (0,0)  {$\dots$};
	\node (5) at (4,0) [rectangle,draw,thick,minimum size=1.2cm] {N};
	\node (4) at (2,0) [circle,draw,thick,minimum size=1.4cm] {N};
	\draw[thick] (1) -- (2) -- (3) -- (4) -- (5);
	\draw [decorate,decoration={brace,amplitude=15pt,mirror},thick,yshift=-1.5em]
	(-2.8,0) -- (2.8,0) node[midway,yshift=-2.5em]{P-1};
	\end{tikzpicture}
\end{center}
The values of the Fourier coefficients $a_k$ and potential $\hat{V}(\sigma,\eta)$ are calculated using eq.(\ref{eq:fourier_vhat}). We find
\begin{eqnarray}
& & a_k= \frac{N P^2}{k^3 \pi^3} \sin\left( \frac{k\pi }{P}\right) \left(1+(-1)^{k+1}\right),\label{ak2}\\
& & 
\hat{V}= \frac{N P^2}{2 \pi ^3} \text{Re}\big(\text{Li}_3(e^{-\frac{\pi}{P}  (| \sigma |-i\eta +i)}) -\text{Li}_3(-e^{-\frac{\pi}{P}  (| \sigma |-i\eta +i)}) +
\text{Li}_3(-e^{-\frac{\pi}{P}  (| \sigma |-i\eta -i)})-\text{Li}_3(e^{-\frac{\pi}{P}  (| \sigma |-i\eta -i)}) \big) \, .\nonumber 
\end{eqnarray}
The holographic central charge is
\begin{eqnarray}
& & c_{hol}= \frac{N^2 P^4}{32\pi^{10}}\left(31\zeta(5) + 8{\text{Li}_5(-e^{\frac{2\pi i}{P}} ) +8 \text{Li}_5(-e^{-\frac{2\pi i}{P}})}  -   8{\text{Li}_5(e^{\frac{2\pi i}{P}} ) -8\text{Li}_5(e^{-\frac{2\pi i}{P}})} \right).\nonumber\\
& & 
c_{hol}\sim \frac{7 N^2 P^2}{4\pi^8}\zeta(3) +O\left(\frac{\log P}{P^2}\right).\label{example2}
\end{eqnarray}

\underline{\bf Central Charge Along the Flow}
\\
Now, let us compute the central charge 
 for the solution in eq.(\ref{tendconfig}), dual to a $5$ dimensional CFT compactified on a circle flowing to a gapped QFT.
 
 We set $d=3$ in eq.(\ref{chol}). Note that this treats the QFT as if it were (3+1) dimensional. Comparing eqs.(\ref{tendconfig}) and (\ref{assignation}), we find,
 \begin{eqnarray}
& &\alpha(r)= f_1(\sigma,\eta)  \frac{2\tilde{g}^2}{9} H^{1/2}(r)r^2,\;\;\;\;\beta(r)=\frac{1}{r^2 f(r)}, \;\;\;\;\
\\
& & ds_{\text{int}}^2= g_{ij}(r,\vec{\theta}) d \theta^i d \theta^j=f_1 \left[\frac{2\tilde{g}^2}{9} H^{-3/2}\, f(r)\, d\phi^2  + f_2 \left(d \theta^2 +\sin{\theta}^2(d \varphi-A_1^{(3)})^2\right)+ f_3 \left(d \sigma^2+ d \eta^2 \right)\right]  , \nonumber\\
& &  V_{\text{int} }= {\cal N} r^3\sqrt{f(r)}, \;\;\;\;\ {\cal N}=  \int d \phi ~d\theta~d\varphi~ d\sigma~ d \eta~(\sin\theta~ f_1^4 f_2 f_3 f_6),\nonumber
\end{eqnarray}
Computing  ${\cal N}$ explicitly we find,
\begin{equation}
{\cal N}=(\frac{2\tilde{g}^2}{9})^2 3^3 \pi^5 L _\phi\int_{0}^{P} d \eta \int_{-\infty}^{\infty} d \sigma~\sigma^3 \partial_\sigma V \partial_\eta^2 V .\label{calN}
\end{equation}
This leads to the  expression for the holographic central charge,
\begin{equation}
c_{hol}= \frac{\cal N}{8 G_N} \frac{(f(r)^2 r^3)}{(f(r)+r/6f'(r))^3} \label{cholr}
\end{equation}
where $G_N= 8\pi^6$. 
In Figure \ref{fig:chol1} we have an example plot for the holographic central charge $c_{hol}$ as a function of the radial coordinate. 
\begin{figure}[h]
    \centering
    \includegraphics[scale=1]{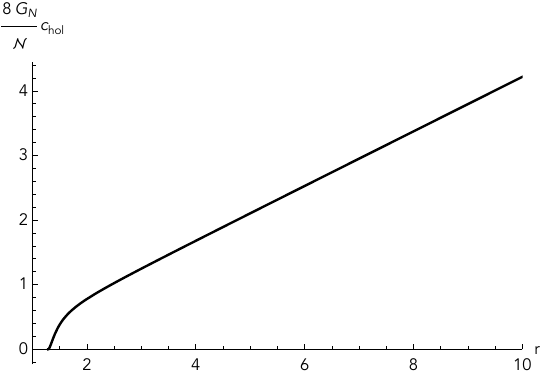}
    \caption{The holographic central charge with parameters $c=1, \mu=1$}
    \label{fig:chol1}
\end{figure}

In the limits of eq.(\ref{cholr}) for $r\rightarrow\infty$ and $r\rightarrow r^*$, we find
\[ \frac{8 G_N}{{\cal N}}c_{hol} = \begin{cases} 
        \frac{243}{128 \tilde g^2}r + O(1/r) &\;\;\;\;\;\;\; r\rightarrow \infty \\
         \frac{ 1944 {r^*}^7 (-12 c^2 \tilde g^2 {r^*}^2 + 12 \tilde g^2 {r^*}^5 - 9 \mu)^2 }{(4 c^4 \tilde g^2 - 20 c^2 \tilde g^2 {r^*}^3 + 16 \tilde g^2 {r^*}^6 - 
  27 {r^*} \mu)^3}(r - 
   {r^*})^2 + O(r - 
   {r^*})^3 &\;\;\;\;\;\;\; r\rightarrow r^*
       \end{cases}
    \]
 This quantity is divergent at large energies, pointing to a UV completion in terms of a 5-dimensional system. In other words, the infinite tower of Kaluza-Klein modes for the $S^1_\phi$ compactification of the QFT are seen by this quantity as a divergent number of degrees of freedom as we explore higher energies. Also, the central charge in eq.(\ref{cholr}) vanishes for $f(r^*)=0$, signalling a gapped system.

Note that the factor ${\cal N}$ defined in eq.(\ref{calN}) contains, aside from constants, the same integral found in the strictly conformal case--see eq. (\ref{vavar}).
The physical intuition behind this is that the degrees of freedom of the 5d UV conformal theory
proportional to the quantity ${\cal N}$ in eq.(\ref{vavar}) get `weighted' by the volume of the compactified manifold. This matches with the picture advocated in \cite{Bobev:2017uzs}--see also \cite{Faedo:2019cvr}. In other words, the free energy of the system along the flow has a contribution coming from the flow itself (the $r$-dependent part of $c_{hol}$) and a contribution coming from the UV SCFT (the factor ${\cal N}$) in eq.(\ref{cholr}).

Due to compactification on the circle, we choose $d=3$ in eq.(\ref{cholr}), instead of $d=4$. The central charge at the UV limit is  linearly divergent. The four-dimensional observable does not capture the five-dimensional conformal UV completion.

The divergence of the free energy towards the UV is expected. Massive fields that originate from the Kaluza-Klein compactification on the circle $S^1_\phi$ generically have a mass inversely proportional to the size of the circle and are frozen at lower energies. When flowing towards the UV, these massive fields are excited, the number of these KK modes grows with energy and causes the divergence of the central charge. 
 A similar phenomena was also observed and explained in \cite{GonzalezLezcano:2022mcd, Merrikin:2022yho}.

 In the next section,  we calculate a different quantity called the `flow central charge', which remedies  this deficiency and is sensitive to the existence of both the IR-gapped QFT and the UV fixed points. It is also monotonic, so it can be used as a measure of the number of degrees of freedom (density of states) for the flow across dimensions.
\subsection{The flow-central charge: a monotonic c-function for the flow across dimensions}\label{flowcentral}
For the case in which the  QFT presents a flow across dimensions (a QFT anisotropic in space),
 we consider a more elaborated definition of the central charge. This was  presented in \cite{Bea:2015fja}.
Consider the metric and dilaton of the form,
\begin{eqnarray}
& & ds^2= -\alpha_0 dt^2+ \alpha_1 dy_1^2 +\alpha_2 dy_2^2 + ....+ \alpha_d dy_d^2+
 \Pi_{i=1}^d (\alpha_1....\alpha_d)^{\frac{ 1 }{ d } } b(r) dr^2+\nonumber\\
 & & g_{ij} (d\theta^ i- A_1^i) (d\theta^j-A_1^j),\;\;\;\;\;\;\;\;\Phi(r,\vec \theta).
\label{eq4x}
\end{eqnarray}
For the case studied here, we set $d=4$ to deal with a five dimensional $(t,x_1,x_2,x_3,\phi)$ anisotropic system. We intend to define a quantity that is monotonous along the flow and that detects the fixed point in the UV limit together with the gapped character of the IR.
Following \cite{Bea:2015fja},
One defines for our background in eq.(\ref{tendconfig}),
\begin{eqnarray}
& & ds^2_{\text{int}}=  \alpha_1 dy_1^2 +\alpha_2 dy_2^2 + ....+ \alpha_d dy_d^2
 +g_{ij} (d\theta^ i- A_1^i) (d\theta^j-A_1^j), \qquad  \Phi(r,\vec \theta).
\label{eq5x}\\
& &  g_{ij} (d\theta^ i- A_1^i) (d\theta^j-A_1^j)=f_1 \left[
    f_2 \left(d \theta^2 +\sin{\theta}^2(d \varphi-A_1^3)^2\right)+ f_3 \left(d \sigma^2+ d \eta^2 \right)\right]  . \nonumber
\end{eqnarray}
We form the combination,
\begin{equation}
V_{\text{int}}= \int_{X} \sqrt{\det[g_{\text{int}}]  e^{-4\Phi}}, \qquad \hat{H}=V_{\text{int}}^2.\label{eq55x}
\end{equation}
The integral is over the manifold $X$ consisting of the internal space $g_{ij}$.
The holographic central charge along the flow--called $c_{\text{flow}}$ below, is
\begin{equation}
c_{\text{flow}}= \frac{d^d}{G_N} b(r)^{d/2}\frac{\hat{H}^{\frac{2d+1}{d}}}{\hat{H}'^d}.\label{centralflowxx}
\end{equation}
Using $d=4$ to work with  five-dimensional anisotropic QFT one has,
\begin{eqnarray}
& & \alpha_0=\alpha_1=\alpha_2=\alpha_3= \frac{2\tilde{g}^2}{9} f_1(\sigma,\eta) H(r)^{1/2}r^2,\;\;\; \alpha_4= \frac{2\tilde{g}^2}{9} f_1(\sigma,\eta) \frac{f(r)}{H(r)^{3/2}},\nonumber\\
& & b(r)= \frac{H(r)^{1/2}}{f(r)^{5/4}r^{3/2}},\;\;\; V_{\text{int}}= {\cal N} r^3 \sqrt{f(r)}  ,~~ \hat{H}={\cal N}^2 r^6 f(r).\;\;\;\nonumber\\ 
& & c_{\text{flow}}=\left(\frac{2}{3} \right)^4 \frac{{\cal N}}{G_N} \frac{H(r) r^4 f(r)^2}{\left(f(r) + \frac{r f'(r)}{6}\right)^4}
,\label{cholflow4}
\end{eqnarray}
with $\cal N$ given in eq.(\ref{calN}). In Figure \ref{fig:cflow} we have an example plot for the flow-central charge. 
\begin{figure}[h]
    \centering
    \includegraphics[scale=1]{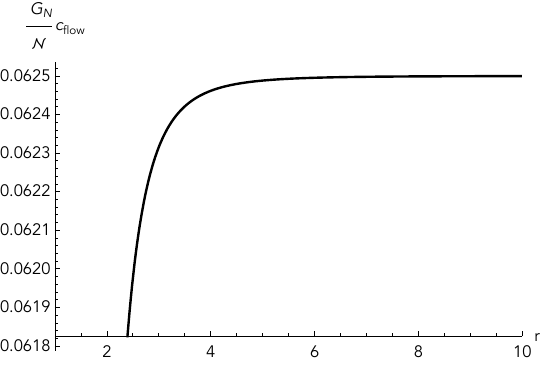}
    \caption{The flow-central charge with $c=1, \mu=1$}
    \label{fig:cflow}
\end{figure}

We find that the flow central charge is {\it monotonic}. In SUSY cases studied previously \cite{Legramandi:2021aqv, Bea:2015fja}, this quantity was shown to be monotonic using the BPS equations. For our background, although it is non-BPS, this is still the case. 
The quantity $c_{\text{flow}}$ properly detects the UV fixed point and the IR gapped character of the theory by showing zero degrees of freedom at $r=r^*$.

We now study a different observable of the QFT, Wilson loops.
\subsection{Wilson Loops}
In this section, we calculate Wilson loop expectation values in QFT to test the proposed  mass gap and confinement/screening at low energies. We start giving a summary of the general formalism to compute Wilson loops in holographic QFTs, following the methods of \cite{Sonnenschein:1999if, Nunez:2009da}. The generic methods are relevant for the study of other probes that reduce to an effective action in the background, like Entanglement Entropy or 't Hooft loops. See also \cite{Nunez:2023nnl}.

For a background of the form 
    \begin{equation}
        ds^{2}=-g_{tt}dt^{2}+g_{xx}d\vec{x}^{2}+g_{rr }
            dr^{2}+g_{ij}d\theta ^{i}d\theta ^{j}\ .\label{wilsonb}
    \end{equation}
With $g_{tt}$, $g_{xx}$, $g_{rr }$ depending only on the radial coordinate $r$, we propose a string embedding, which has a Nambu-Goto action of the form,
\begin{eqnarray}
& & t=\tau,~~~x=x(\gamma),~~~r=r(\gamma).\nonumber\\
& & S_{NG}= T_{F1} \int d\tau d\gamma \sqrt{g_{tt}(r) g_{xx}(r)x'^2 + g_{tt}(r)g_{rr}(r) r'^2}.\label{NGwilson}
\end{eqnarray}
The embedding is parameterised in terms of $(\tau,\gamma)$ chosen as the worldsheet coordinates.
Following \cite{Nunez:2009da}, the equations of motion for the string moving in this background read
    \begin{equation}
        \frac{dr }{d\gamma }=\pm \frac{dx}{d\gamma }V_{eff}\left( r \right) \, .\label{eqmov}
    \end{equation}
The `effective potential' is defined in \cite{Nunez:2009da} as,
    \begin{equation}
        V_{eff}\left( r \right) =\frac{F\left( r \right) }{CG\left( r
            \right) }\sqrt{F^{2}\left( r \right) -C^{2}}\ ,~~ F^2\left( r \right) =g_{tt}g_{xx},~~G^2\left( r \right)=g_{tt}g_{rr},\label{potdef}
    \end{equation}
     in which the constant $C=\frac{F^2x'}{\sqrt{F^2 x'^2+ G^2 r'^2}}$ can be obtained from the equations of motion. 
     
     If one fixes the coordinate as $x(\gamma)=\gamma$ one can find eq.(\ref{eqmov}) from the conserved `Hamiltonian' with the relation $C= F(r_0)$. Here, $r_0$ is the turning point of the string embedded in the background, satisfying $r'(\gamma)=0$. We fix the gauge in this way in what follows and choose $C=F(r_0)$ from here on. 

We observe that in this formalism, we have  an open string whose endpoints are at a D-brane, placed at $r \to \infty$. Dirichlet boundary conditions  are chosen for the string at $r \rightarrow \infty $ by imposing $V_{eff}|_{r \rightarrow \infty}\sim \infty $
The separation between the two ends of the string can be thought of as the separation between a quark-antiquark pair. The energy between the pair of quarks is calculated from the Nambu-Goto action. There is a regularisation procedure implemented by subtracting the energy of two non-dynamical strings extended along the whole range of the radial coordinate $[r^*, \infty)$, which subtracts the rest mass of the quark-antiquark pair. 

The string takes a U-shape in the bulk. The separation and energy can be written in terms of the distance from the position of the turning point of the string $r_0$, by
    \begin{align}
        L_{QQ}\left( r_{0}\right) &=2\int_{r_{0}}^{+\infty }
        \frac{dz}{V_{eff}(z) }\, , \label{QQ separation}  \\
    E_{QQ}\left( r_{0}\right) &=F\left( r_{0}\right) L_{QQ}\left( r_{0}\right)
            +2\int_{r_{0}}^{+\infty }dz\frac{G\left( z\right) }{F\left( z\right) }
            \sqrt{F\left( z\right) ^{2}-F\left( r_{0}\right) ^{2}}-2\int_{r^*}^{+\infty }dz\
            G\left( z\right) \label{QQ energy} \, .
    \end{align}

Conditions for confinement or screening, finite or infinite quark separation, are given in \cite{Nunez:2009da}.
Now we will apply the mentioned general treatment to our background in eq.(\ref{defi}).
Assuming the string embedding in the $\Sigma$ plane to set at constant value $(\sigma,\eta)=(\sigma^*,\eta^*)$, we pick coordinates $t=\tau, x=\gamma, r=r(\gamma)$. One has
\begin{equation}
ds^2_{\text{ind}}= \frac{2\tilde{g}^2}{9} f_1(r,\sigma^*,\eta^*)\left(-H(r)^{1/2}r^2 d \tau^2+H(r)^{1/2}r^2 \left(1+\frac{r'(\gamma)^2}{r^2 f(r)}\right)d \gamma^2\right),
\end{equation}
\begin{equation}
S_{NG}= T_{F1}\int d\tau d\gamma \sqrt{\det[g_{\alpha \beta}]}, =T_{F1} T \frac{2\tilde{g}^2}{9} \int d\gamma \sqrt{F^2+G^2r'^2}, \label{NGwilson2}
\end{equation}
with function definitions
\begin{equation}
F(r)=\sqrt{H(r)}f_1(r,\sigma^*,\eta^*)r^2 \qquad G(r)=\sqrt{\frac{H(r)}{f(r)}}f_1(r,\sigma^*,\eta^*)r.
\end{equation}
For the effective potential one has
\begin{equation}
V_{eff}=\frac{\sqrt{f(r)}r}{f_1(r_0,\sigma^*,\eta^*)\sqrt{H(r_0)}r_0^2} \sqrt{f_1(r,\sigma^*,\eta^*)^2H(r)r^4-f_1(r_0,\sigma^*,\eta^*)^2H(r_0)r_0^4}.
\end{equation}
By expanding effective potential close to $r_0=r^*$, $V_{eff}\sim (r-r^*)$, indicating that $L_{QQ}$ diverges as $r_0$ approaches $r^*$, see the paper \cite{Nunez:2009da}. This can be taken as a possible indication of the confining behaviour in our dual QFT. We analyse this below in more detail.

To investigate the low energy behaviour of the QFT and to check for confinement (or screening), we analyse the expressions for the length and the energy of the quarks pair.
From eqs.(\ref{QQ separation})-(\ref{QQ energy}), we read the length of the quark-antiquark pair and its energy to be
\begin{eqnarray}
& & L_{QQ}\left( r_{0}\right) = f_1(r_0,\sigma^*,\eta^*)\sqrt{H(r_0)}r_0^2 \times \nonumber\\
 & &  \int_{r_{0}}^{\infty }\frac{1}{\sqrt{f_1(r,\sigma^*,\eta^*)^2H(r)r^4-f_1(r_0,\sigma^*,\eta^*)^2H(r_0)r_0^4}\sqrt{f(r)}r} dr\ ,\label{Lqq} \\
& & E_{QQ} \left( r_{0}\right) =F(r_{0}) L_{QQ} \left( r_{0}\right) +
{2} \int_{r_{0} }^{\infty } dr \frac{\sqrt{f_1(r,\sigma^*,\eta^*)^2H(r)r^4-f_1(r_0,\sigma^*,\eta^*)^2H(r_0)r_0^4}}{\sqrt{f(r)}r} -\nonumber\\
 & & 
{2} \int_{r^* }^{\infty } dr \sqrt{\frac{H(r)}{f(r)}}f_1(r,\sigma^*,\eta^*)r
 \ .\label{Eqq}
\end{eqnarray}%
The integrals can not be performed analytically and are calculated with numerical methods.

In Figure \ref{fig:EQQ}, a plot for the length of the string between the quark anti-quark pair and the energy as a function of this length for certain parameter values is provided. 
\begin{figure}[h]
    \centering
    \includegraphics[scale=0.7]{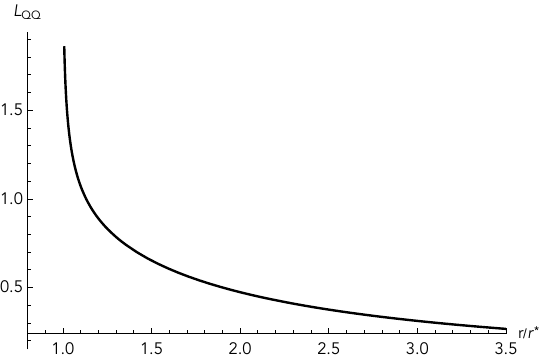}
    \hspace{1cm}
    \includegraphics[scale=0.65]{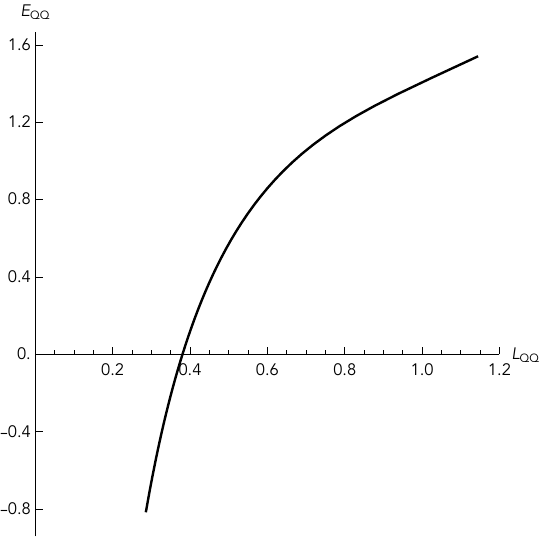}
    \caption{The length of the string between quark anti-quark pair and the energy with $c=1, \mu=1$}
    \label{fig:EQQ}
\end{figure}
The resulting configurations for the embedding of the string with $c=1, \mu=1$ with different separation lengths are presented in Figure \ref{fig:Wline}. These show the conventional behaviour of the confining QFTs. We have chosen $\sigma^*=0$ and $\eta^*$ fixed. Using identities in Appendix \ref{usefulidentities}, we see that the function  $f_1(r,\sigma,\eta)$ is $r$-independent, hence there are significant simplifications in the calculation. The value of $\eta^*$ is directly related to the gauge node with which the Wilson is associated (See \cite{Fatemiabhari:2022kpv} and references therein).
\begin{figure}
    \centering
    \includegraphics[scale=1.2]{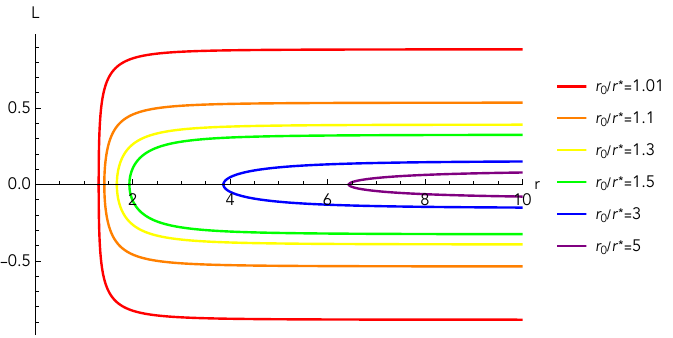}
    \caption{Embedded string profile for Wilson loop calculations with $c=1, \mu=1$}
    \label{fig:Wline}
    \end{figure}
    
Note that the concavity of the energy curve $E_{QQ}$ as a function of $L_{QQ}$ is downwards, indicating  the stability of the embedded probe string  configuration of eq.(\ref{NGwilson}). Note also that for the large $L_{QQ}$, the energy grows linearly, suggesting a confining behaviour. Below, we discuss a phenomenon not captured by the string configuration studied in this section. The existence of this configuration is related to the presence of sources (flavour branes) in this family of backgrounds.
\subsubsection{{Screening}}\label{screensec}
Massless flavour quarks are present in our dual QFT setups. In fact, for every kink in the convex, piece-wise linear rank function ${\cal R}(\eta)$, we encounter a flavour group, a set of D7 branes localised in the $\eta$-direction, as indicated in eq.(\ref{chargesfinal}).

The presence of these favour groups makes the screening phenomenon possible. This can indeed happen by creating a pair of flavour quarks that break the flux tube between the inserted heavy probe quark and the anti-quark pair. 

Even if the Wilson loop is associated with a gauge node without flavour groups attached, there can be a chain of interactions across the quiver that finally excites the flavour quarks. Studying this proposal rigorously requires the insertion of a probe string in bulk, which has the possibility to extend not only in spatial $x_1$ and r directions but also in $\eta$ direction, being related to the gauge node in the quiver.

In general, the configuration is parameterised in terms of $(\tau,\gamma)$. The induced metric reads
\begin{eqnarray}
& & t=\tau,~~x=x(\gamma),~~r=r(\gamma),~~\eta=\eta(\gamma).\label{parametri}\\
& & ds_{ind}^2=\left( \frac{2\tilde{g}^2}{9}\right) f_1(r,\sigma^*,\eta)\Big[- H(r)^{1/2} r^2 d\tau^2 + d\gamma^2\left( H(r)^{1/2} r^2 x'^2 +\frac{H(r)^{1/2}}{f(r)} r'^2+ \frac{9 f_3(r,\sigma^*,\eta) }{2\tilde{g}^2}\eta'^2\right)  \Big].\nonumber
\end{eqnarray}
After integration over $0\leq\tau\leq T$, the Nambu-Goto action reads,
\begin{eqnarray}
& &S_{NG}= T T_{F1}\int d\gamma \sqrt{F^2 x'^2+ G^2 r'^2+ S^2 \eta'^2},\label{NGgeneric}\\
& & 
F^2=\left(\frac{2\tilde{g}^2}{9}\right)^2 f_1^2(r,\sigma^*,\eta)H(r) r^4,~~ ~~~~
G^2=\left(\frac{2\tilde{g}^2}{9}\right)^2 f_1^2(r,\sigma^*,\eta)\frac{H(r) r^2}{f(r)},\nonumber\\
& & S^2=\left(\frac{2\tilde{g}}{9}\right) f_1^2(r,\sigma^*,\eta) f_3(r,\sigma^*,\eta) H(r)^{1/2} r^2.\nonumber
\end{eqnarray}
This action should be minimised to learn if a configuration extending on the $\eta$-direction can reach the closest flavour group. This is a complicated problem that we leave for future study. 

Here, we pursue a simpler analysis that captures the essence of the calculation. We insert a probe string picking $t=\tau, \eta=\gamma$ with other coordinates fixed. We calculate the Nambu-Goto action for this embedding, assuming that the string goes from a chosen node at $\eta=\eta^*$ to some other node containing flavour degrees of freedom. 

If the energy of this configuration is lower than the conventional Wilson loop expectation value calculated in the previous section, the new solution takes over, and there is a phase transition to this new solution. This predicts the breaking of the flux tube when the separation between the probe quark anti-quark pair implies an energy cost that is enough to create the dynamical quark pair. 

We then consider a configuration with $x=x_0$, $r=\bar{r}$ and $\sigma^*=0$. For this configuration, we find
\begin{eqnarray}
& & ds^2_{\text{ind}}=- \frac{2\tilde{g}^2}{9} f_1(\bar r,\sigma^*,\eta)H(\bar r)^{1/2}\bar r^2 d \t^2+f_1(\bar r,\sigma^*,\eta)f_3(\bar r,\sigma^*,\eta)d \eta^2,\\
& & S_{NG}= T_{F1}\int d\tau d\gamma \sqrt{\det[g_{\alpha \beta}]}, =T_{F1} T \int d\eta \sqrt{S^2}, \label{NGwilson3}
\end{eqnarray}
We choose $\sigma^*=0$ and using the relations in Appendix \ref{usefulidentities}, one has the simplified results
\begin{eqnarray}
& &     f_1^2(\bar r,\sigma^*,\eta)= \frac{9\pi^2}{4 X^4} \left( \frac{3 X^4 \sigma \partial_\sigma V +\sigma^2\partial_\eta^2 V}{\partial_{\eta}^2  V}\right), \qquad  f_3(\bar r,\sigma^*,\eta)= \frac{X^2\partial_{\eta}^2  V}{3 \sigma \partial_\sigma V} .\\
& & f_1^2 f_3=\frac{3\pi^2}{4 X^2}\left( \frac{\sigma\partial_\eta^2 V}{\partial_\sigma V} + 3 X^4 \right)|_{\sigma^*=0}\simeq \frac{9\pi^2}{4}X^2(\bar{r}),~~~ S^2\simeq \frac{\pi^2}{2} \tilde{g}^2 \bar{r}^2.\nonumber
\end{eqnarray}

Hence we have
\begin{eqnarray}
    S_{NG}= T_{F1} T \frac{\pi\tilde{g}}{\sqrt{2}} \bar r \int d\eta ,
\end{eqnarray}
which minimises when $\bar r = r^*$. The integration in $\eta$ is from $\eta^*$ (the gauge group for which we compute the Wilson loop) to the desired $\eta=\eta_F$ position with flavour branes. Hence, there are fixed energy configurations that take over and are the preferred configuration relative to the embedding presented in the previous subsection. We interpret this as screening due to the presence of the dynamical fundamental fields.

In Appendix \ref{appendixW} we consider a string that stretches in the $\eta$-direction as $x$ is increased. We find a phase transition when the turn-around position $\eta_0$
approaches the location of the flavour group.
The connected $U$-type configuration in $\eta$ has a finite length. This analysis should be extended to the string embedded in the $r,\eta$ directions. We leave this for future work.
\subsection{Entanglement Entropy}
For QFTs with a string theory dual, the Entanglement Entropy (EE) between two regions can be calculated with the prescription presented in \cite{Ryu:2006bv,Ryu:2006ef}. The method is based on finding a minimal  eight-dimensional surface in the gravity side
such that the boundary of this surface is coincident with the boundary of the two entangled regions. We divide the space into two regions, one of them a region in the shape of a strip of length $L_{EE}$ and the other its complement.
The entanglement entropy between these two regions is calculated in \cite{Ryu:2006bv}-\cite{Klebanov:2007ws} by minimising 
\begin{equation}
S_{EE}=\frac{1}{4 G_N}\int_{\Sigma_8} d^8\sigma \sqrt{e^{-4\Phi} \det[g_{ind}]}.\label{SEEdef}
\end{equation}

There can be different eight surfaces that  minimise $S_{EE}$ in eq.(\ref{SEEdef}). Hence, there is a possibility for phase transitions among different extremal surfaces. 
In \cite{Klebanov:2007ws}, it was proposed that the presence of a phase transition in the EE can be a criterion for confinement. This proposal was critically analysed in  more detail in \cite{Kol:2014nqa,Jokela:2020wgs}. In these papers, it was found that phase transitions can be absent in certain confining models and that non-confining models can display a phase transition.

Following \cite{Ryu:2006bv,Ryu:2006ef}, and keeping in mind the generalised treatment of  \cite{Klebanov:2007ws,Kol:2014nqa}, we calculate the entanglement entropy of a strip-like region of the QFT. This is accomplished by computing the area of the eight-surface $[x_1,x_2,x_3,\phi, \theta
,\varphi,\sigma,\eta]$ with $r=r(x_1)$ in the background of Section \ref{sec:background}. 
The induced metric on the eight-surface and the entanglement entropy are,
\begin{eqnarray}
ds^{2}_{st} &=&f_1 \frac{2\tilde{g}^2}{9} \Big[ H(r)^{1/2}\, r^2\big(~ (1+\frac{r'^2}{r^2 f(r)})\, dx_1^2 + \,
dx_2^2+dx_3^2 ~\big)+ H(r)^{-3/2}\, f(r)\, d\phi^2\Big] \;\;\;\\ \nonumber
& &+f_1f_2 d \tilde\Omega_2+f_1f_3 (d\sigma^2+d \eta^2).\;\;\;\\
& &\sqrt{e^{-4\Phi}\text{det}[g_8]}=\sqrt{f_1^8f_2^2f_3^2f_6^2 \sin^2\theta \frac{2\tilde{g}^2}{9}^4 f(r) r^6(1+\frac{r'^2}{r^2 f(r)})},\nonumber\\
 S_{EE}&=& \frac{1}{4 G_N}\int d^8x \sqrt{e^{-4\Phi}\det[g_8]}=\mathcal{\hat N} \int_{-L}^{L}dx\sqrt{r^6 f(r)(1+\frac{r'^2}{r^2 f(r)})}.\label{EE211}
\end{eqnarray}
We defined,
\begin{eqnarray}
& &\mathcal{\hat N} = \frac{3^3\pi^5}{4 G_N}L_yL_z L_\phi{(\frac{2 \tilde{g}^2}{9} )^2} \int_{0}^{P} d \eta \int_{-\infty}^{\infty} d \sigma~\sigma^3 \partial_\sigma V \partial_\eta^2 V.\label{messi}
\end{eqnarray}
From eq.(\ref{potdef}), we read 
\begin{equation}
F(r)= r^3\sqrt{f(r)},~~~~G(r)=r^2.\label{FGEE}
\end{equation}
Following the conserved Hamiltonian technique, as in the Wilson loop calculation summarised in eqs.(\ref{wilsonb})-(\ref{QQ energy}), we minimise the $S_{EE}$. The entanglement entropy must be regularised by subtracting the volume of the two eight-surfaces starting from infinity and ending at the end of the space $r^*$. Then, we find  the length of the interval and the entanglement entropy by computation of the regulated area of the surface that turns around at $r_0$,
\begin{align}
& L_{EE}\left(r_0\right)=2 r_0^3\sqrt{f(r_0)}\int_{r_0}^{\infty} \frac{d r}{r\sqrt{f(r)}\sqrt{r^6f(r)-r_0^6f(r_0)}}, \\
& S_{EE}\left(r_0\right)= \mathcal{\hat N} \int_{r_0}^{\infty} \frac{r^5\sqrt{f(r)}}{\sqrt{r^6f(r)-r_0^6f(r_0)}} d r-\mathcal{\hat N}  \int_{r^*}^{\infty} r^2 d r .\label{SEE}
\end{align}

These integrals can not be performed analytically, so we rely on numeric calculations. 
The resulting plot for $c=1, \mu=1$ is presented in Figure \ref{fig:SEE}.
\begin{figure}
    \centering
    \includegraphics[scale=0.8]{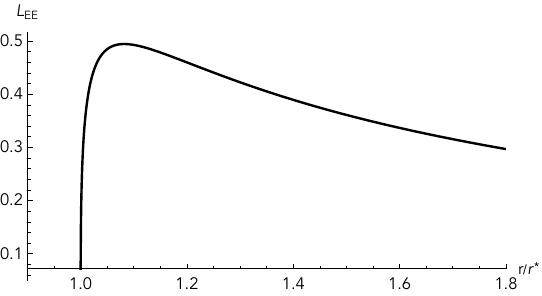}
    \hspace{1cm}
    \includegraphics[scale=0.8]{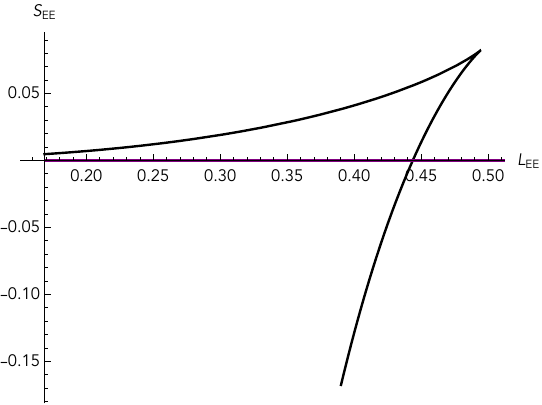}
    \caption{$L_{EE}$ and entanglement entropy respective to $L$ with $c=1, \mu=1$}
    \label{fig:SEE}
\end{figure}
From the left panel in Figure \ref{fig:SEE}, one realizes that $L_{EE}$ is not a monotonous function, but actually going from a vanishing value at $r_0=r^*$ to asymptotically zero at $r_0\to\infty$. This behaviour generates the possibility of a phase transition.
Indeed, the conditions for the emergence of a phase transition according to \cite{Kol:2014nqa} 
are satisfied. More specifically equations (2.26)-(2.29) of \cite{Kol:2014nqa} imply $j=3$ permitting a phase transition. 

One can also check that by setting $c=0, \mu=0$, which takes our solution back to $AdS_6$ solution,  $L_{EE}$ does have a monotonic behaviour, and the phase transition disappears. In this case, the integrals in eqs.(\ref{SEE}) can be performed explicitly, finding
$L_{EE}\sim \frac{1}{r_0}$ and $S_{EE}\sim \frac{\hat{\cal N}}{L_{EE}^3}$.

The right panel of Figure \ref{fig:SEE} for $S_{EE}(r_0)$ in terms of $L_{EE}$  shows a downwards concavity indicating  that the configuration is stable. This plot also includes the disconnected solution, which is renormalised to zero. After the phase transition, the disconnected solution is preferred by the system. 
In view of the content of Section \ref{screensec}, one should consider more involved eight-surfaces than the one considered below eq.(\ref{SEEdef}), along the lines of allowing, for example $r(x_1,\eta)$. We leave this for future work.

\subsection{Holographic Complexity}
We start with a very brief review of the concept of complexity in QFTs. Then, following \cite{Reynolds:2017lwq}, we present a proposal for the calculation of this quantity holographically, using the backgrounds of eq.(\ref{tendconfig}). We focus on the proposal known as the CV conjecture.

The quantum computational complexity is defined as a measure of the minimum number of elementary gates one needs in a quantum circuit to construct a generic state in the Hilbert space starting from a specified reference state. 
In \cite{Susskind:2014rva}, Susskind conjectured a relation between the bulk geometry and the dual boundary state, proposing that  the complexity of the dual boundary state can be related to the time-dependent geometry of the region behind the horizon of an AdS black hole.

The conjecture posited in \cite{Stanford:2014jda} refines the previous idea by suggesting that the computational complexity of the boundary state at a specific time (on a designated spacelike slice of the boundary) is correlated with the volume of a maximal spacelike slice in the bulk, ending on the specified boundary slice. This is referred to as the CV conjecture. 

In this line of thought, one specifies some spatial slice $\Sigma$ on the boundary of the
spacetime. The complexity $\mathcal C$ of a pure state $|\Psi \rangle$ of a holographic field theory on this patch will be given by the volume of a  co-dimension one slice $B$ in bulk. This submanifold in the bulk, whose volume obeys some maximal condition, has its boundary on $\Sigma$.
\begin{equation}
{\mathcal C_{\textnormal{V}}} \propto \frac{V(B)}{G_{\textnormal{N}} l_{\textnormal{AdS}}}.
\end{equation}
%
 For discussions about divergences appearing in the holographic complexity calculation, the reader is referred to \cite{Reynolds:2017lwq}.
 
A simple example of the calculation is to consider vacuum AdS$_{6}$ in Poincar\'e coordinates,
\begin{equation} \label{adsp}
ds^2 = \frac{\ell^2}{r^2} dr^2 + \frac{r^2}{\ell^2}\left(- dt^2 + d\vec{x}_4^2\right),
\end{equation}
which is dual to a CFT in a flat space.  
The maximal volume slice with boundary at fixed time $t=0$ is the surface in the bulk with volume 
\begin{equation} \label{vol}
V(B) = \int \mathop{dr}\mathop{d^{4}x} \sqrt{h} = \frac{V_{\vec x}}{\ell^3}  \int_0^{r_M} r^3 {\mathop{dr}} = \frac{ V_{\vec x}}{4 \ell^3} r_M^4,
\end{equation}
where $V_{\vec x}$ is the coordinate volume in the $\vec x_4$ non compact directions which reflects an IR divergence. Thus, the complexity calculated following  the CV prescription is, 
\begin{equation}  \label{cvol}
\mathcal{C}_{\textnormal{V}} \propto  \frac{  V_{\vec x}}{4 G_6 \ell^4} r_M^4.
\end{equation}
This result is proportional to the volume of the space on which the field theory is formulated and proportional to the power of the UV-cutoff, reflecting a UV divergence. 

We can lift the AdS$_6$ background to type IIB, as explained in Appendix \ref{appendix1}, see also  \cite{Legramandi:2021aqv} and \cite{Hong:2018amk}.
The lifted type IIB background is given by that in eq.(\ref{tendconfig}), in the special case $X(r)=1$, $H(r)=1$, $A_1^{(3)}=0$ and the parameters $c=\mu=0$.
The constant time-slice has a nine-dimensional line element and dilaton given by,
\begin{eqnarray}
& & ds_9^2= f_1 \Big[ \left(\frac{2\tilde{g}^2}{9}  \right)  r^2 d\vec{x}_3^2+  \left(\frac{2\tilde{g}^2}{9}  \right)^2  r^2d\phi^2  + \frac{ dr^2}{ r^2} + f_2 d\Omega_2 + f_3 (d\eta^2+d\sigma^2) \Big],~~~ e^{-4\Phi}= f_6^2.\label{9-manifold}\nonumber
\end{eqnarray}
\\
\underline{\bf A proposal to calculate the complexity}
\\
We now {\it propose} a way to calculate the complexity in this case (notice that we are in the case of zero temperature). We  define the complexity as the volume of the nine-manifold weighted by the dilaton and the global conformal factor ${\cal A}$ ( with ${\cal A}=f_1$ in this case), according to
\begin{equation}
\mathcal{C}_V \propto \frac{1}{G_{N,10}}\int dx^9 \sqrt{\frac{e^{-4\Phi}\text{det}[g_9]}{\mathcal{A}}}.\label{julian}
\end{equation}
Calculating explicitly we find
\begin{eqnarray}
 & & \int dx^9 \sqrt{\frac{e^{-4\Phi}\text{det}[g_9]}{\mathcal{A}}}
 = 4\pi  V_{\vec{x}} \left( \frac{2\tilde{g}^2}{9}\right)^{5/2} \left[ \int d\sigma d\eta f_1^4 f_2f_3f_6\right] \int_0^{r_M} r^3 dr,\nonumber\\
 & & \mathcal{C}_V \propto\frac{3^3\pi^5}{G_{N,10}}\left( \frac{2\tilde{g}^2}{9}\right)^{5/2}V_{\vec{x}} \Big(\int d\sigma d\eta \sigma^3 \partial_\sigma V \partial_\eta^2 V \Big) r_M^4.\label{complexityAdS6}
\end{eqnarray}
Let us analyse this result. First, note that the factor $ \Big(\int d\sigma d\eta \sigma^3 \partial_\sigma V \partial_\eta^2 V \Big)$ appears also in the holographic central charge and the Entanglement Entropy--see eqs.(\ref{vavar}), (\ref{cholflow4}) and (\ref{messi}). This implies that the complexity is proportional to the
number of degrees of freedom of the UV CFT, and depends on the parameters defining the UV quiver (see the examples in Section \ref{choldef}). 

Importantly, this factor combined with the volume of the two sphere and the ten-dimensional Newton constant $G_{N,10}$ produce  the Newton constant in six dimensions $G_6$, in correspondence with eq.(\ref{cvol}). The divergence with the UV-cutoff $r_M$ is the same (as it is the IR divergence associated with $V_{\vec{x}}$), when computed both in the reduced and in the lifted backgrounds. Note that this result heavily relies on the definition in eq.(\ref{julian}), importantly including the quotient by the overall conformal factor ${\cal A}$.

We  test our proposal for the complete background in eq.(\ref{tendconfig})--in this case, with the r-dependent functions $X(r), H(r), A_1^{(3)}$ and the non-vanishing parameters $c,\mu$. This background deforms the CFT$_5$ as we discussed above, into a confining four dimensional QFT at low energies. Applying the prescription in eq.(\ref{julian}) to the metric and dilaton in eq.(\ref{tendconfig}), we find
\begin{align}
e^{-4\Phi}\text{det}[g_9]/\mathcal{A} &= (2\gt^2/9)^5f_1^8 f_2^2f_3^2f_6^2 r^6 \sqrt{H} \nonumber\\
\mathcal{C}_V \propto \frac{1}{G_{N,10}}\int dx^9 \sqrt{e^{-4\Phi}\text{det}[g_9]/\mathcal{A}}&=\frac{3^3\pi^5}{G_{N,10}}(\frac{2\gt^2}{9})^{5/2}~\text{Vol}_{R_3} ~L_\phi
\Big(\int d \sigma d \eta \sigma^3\partial_\sigma V \partial^2_\eta V\Big)\int^{r_M}_{r^*}d r r^3 H^{1/4}, \label{complexity}
\end{align}
where $\text{Vol}_{R_3} $ is the volume for the 3 flat directions $(x_1,x_2,x_3)$.

The above analysis, relating the complexity with the central charge of the UV CFT$_5$ is applicable to this expression. Note also that the UV behaviour in terms of the cutoff $r_M$ is similar to that in eq.(\ref{complexityAdS6}), but the behaviour coming from the IR part of the background is different from the one in the CFT case.
If we computed the complexity in the six dimensional solution that lifts to the background in eq.(\ref{tendconfig})--see Appendix \ref{appendix1} we obtain the same $\int^{r_M}_{r^*}d r r^3 H^{1/4}$ factor.

Note that the complexity in eq.(\ref{complexity}), has a UV contribution coming from the upper end of the integral. This is proportional to $r_M^4$, matching the behaviour encountered in the pure AdS$_6$ case of eq.(\ref{complexityAdS6}). Interestingly, there is a contribution from the IR part of the geometry. This is proportional to a Hypergeometric function evaluated at $r^*$. In this way, the mass gap scale enters in the complexity. This is similar to the findings of \cite{Frey:2023qdv}. It is interesting to compare the complexity calculated in eq.(\ref{complexityAdS6}) with that in eq.(\ref{complexity}) as this gives an idea of the complexity of formation, along the lines of \cite{Yang:2023qxx}. It would be interesting to calculate the other definitions of complexity to perform comparisons as those in \cite{Yang:2023qxx}.

It is worth mentioning that in \cite{Belin:2021bga, Belin:2022xmt}, authors introduced an infinite family of observables in asymptotically AdS space defined on codimension-one slices of the geometry, being equally viable candidates for a dual of complexity. All of these observables display the features desired for the dual of the complexity. Our proposal can be studied further to understand whether it fits in this general picture.

\subsection{Spin-two glueballs}\label{glueballssection}

 For the four-dimensional QFT described holographically by the background in eq.(\ref{tendconfig}), we use the six-dimensional  reduction  in eq.(\ref{6dBackground}) to investigate glueball-like excitations. The glueball excitations are analysed by studying fluctuations of background fields in the 6d action. The equations of motion are highly nonlinear and coupled, implying that any given fluctuation may excite fluctuations for other fields in the background. The dynamics for the small fluctuations are governed by linear and coupled second-order coupled differential equations which are very challenging to solve.

Here we consider a special subset of fluctuations which are easier to deal with. We briefly introduce and summarise the procedure for analysing the special kind of fluctuations following \cite{Bachas:2011xa}. It is useful also to refer to Appendix C in the paper  \cite{Nunez:2023xgl}.

The background metric in the eq.(\ref{6dBackground}) is,
\begin{equation}\label{6dBackgroundfl}
ds_6^2 = H(r)^{1/2}\, \left( r^2\, dx^2_{1,3}+f(r)^{-1}\, dr^2\right) + H(r)^{-3/2}\, f(r)\, d\phi^2  \,
        \end{equation}
with functions $f(r)$ and $H(r)$ defined in eq.(\ref{tendconfig}).
We aim to study metric fluctuations of the form
    \begin{equation}\label{flmetric}
        \delta g_{\mu\nu} = e^{2A} \bar{h}_{\mu\nu}, \quad 
        \bar{h}_{\mu\nu} = \begin{pmatrix}
                        h_{ab}(x)\tilde{\psi}(y)& 0 \\
                         0 & 0
                        \end{pmatrix}. 
    \end{equation}
Here the metric fluctuations are parallel to the flat subspace, $x^{a}$ denoting the flat space-time coordinates $(t,x_1,x_2,x_3)$, while $y^{i}$ corresponding to the directions $(r,\phi)$. We use the transverse-traceless gauge
    \begin{equation}
         h^a_{\phantom{a}a} = 0, \quad \nabla_{a}h^{a}_{\phantom{a}b}=0.
    \end{equation}
    
As shown in \cite{Bachas:2011xa}, it is consistent to take vanishing fluctuations for other fields. From here on, we rewrite the background metric as
    \begin{equation}
        ds^{2} = e^{2A(\vec{y})}\left( ds^{2}(\mathcal{M}_{4}) + \bar{g}_{ab}(\vec{y}) dy^a dy^b \right) .\label{genericmetric}
    \end{equation}
    The $\mathcal{M}_{4}$ is maximally symmetric space $dx_{1,3}^2$, which is the four dimensional Minkowski space.
Comparing eq.(\ref{6dBackgroundfl}) with eq.(\ref{genericmetric}), we find

    \begin{eqnarray}\label{metricaIE}
  & &  e^{2A(\vec{y})}= r^2 \sqrt{H(r)}, \\
        & & \bar{g}_{ab}dy^a dy^b=  \frac{f(r)}{r^2H^2(r)\,} d\phi^{2} 
            + \frac{1}{ r^{2}f(r)}dr^2 . \nonumber
        \end{eqnarray}       
This gives $\sqrt{\det \bar{g}_{ab}} = \frac{1}{r^2 H(r)}$.

By introducing the change of variable $\tilde{\psi}(y) = e^{-2A}\Psi (y)$, as suggested in \cite{Bachas:2011xa}, the new function $\Psi(y)$ solves a Schroedinger-like equation 
  \begin{equation}\label{Schroedinger}
        - \bar{\square}_{y} \Psi + V(y)\Psi =  M^{2}\Psi,
    \end{equation}
with the potential defined as
            \begin{equation}\label{potentialfl}
                V(y) = e^{-2A}\bar{\square}_{y}  e^{2A} = \frac{e^{-2A} }{ \sqrt{\det \bar{g}_{ab} } }\partial_a \left[ \sqrt{\det \bar{g}_{ab} } \bar{g}^{ab}\partial_b e^{2A}\right].
            \end{equation}
            
After applying this formalism to the background in eq.(\ref{6dBackgroundfl}) we focus on the glueballs that depend on the radial direction $r$ and the angular coordinate $\phi$ as, 
    \begin{equation}
        \Psi(r,\phi) = e^{i \frac{2\pi}{L_{\phi}}n \varphi}\Psi(r),~~~n\in \mathbb{Z},
    \end{equation}
with the  potential in eq.(\ref{potentialfl}) as,
    \begin{equation}\label{potsch}
        V(r)= \sqrt{H(r)}\partial_r\left( \frac{f(r)}{H(r)}\partial_r(r^2 \sqrt{H(r)})\right).
    \end{equation}
The equation in \eqref{Schroedinger} reads,
    \begin{equation}\label{schrof}
        - r^2 H(r)\partial_r\left( \frac{f(r)}{H(r)}\partial_r \Psi(r)\right)  + \left(V(r) + \frac{n^2r^2H^2(r)}{f(r)}\left(\frac{2\pi}{L_{\varphi}}\right)^{2} \right)\Psi(r) = M^{2} \Psi(r).
    \end{equation}

Following \cite{Nunez:2023xgl} we move to the tortoise coordinate $d\rho=\frac{dr}{r\sqrt{f(r)}}$ and use the change of variables $\Psi(\rho)= \left(\frac{f(\rho)}{\rho^2H^2(\rho)\,} \right)^{-1/4}\Theta(\rho)$
leading to the equation,
    \begin{align}\label{Schrodinger2}
         - \frac{d^2\Theta}{d\rho^2} &+\tilde{V}(\rho) \Theta = M^2\Theta,\;\\
        \tilde{V}(\rho)&= \left( V(r) +  \frac{n^2r^2H^2(r)}{f(r)}\left(\frac{2\pi}{L_{\varphi}}\right)^{2} + r^2 H(r)\frac{d}{dr} \left[ \frac{f(r)}{H(r)}\frac{d} {dr} \left(\frac{f(\rho)}{\rho^2H^2(\rho)\,} \right)^{-1/4} \right]\right)\bigg|_{r=r(\rho)}.
    \end{align} 
This needs to be evaluated in the new $\r$ coordinate. The potential after simplification is
    \begin{equation}
        \tilde{V}(\r) =\left(\frac{-3 r ^2 f'(r )^2+4 r  f(r ) \left(r  f''(r )+3 f'(r )+4 \left(\frac{2\pi n}{L_{\varphi}}\right)^{2} r ^3 H(r )^2\right)+12 f(r )^2}{16 r ^2 f(r )^2}\right)\bigg|_{r=r(\r)}, 
    \end{equation}

The effective potential plotted for different values of $n$ is provided in Figure \ref{fig:pot}. 

\begin{figure}[h]
    \centering
    \includegraphics[scale=0.55]{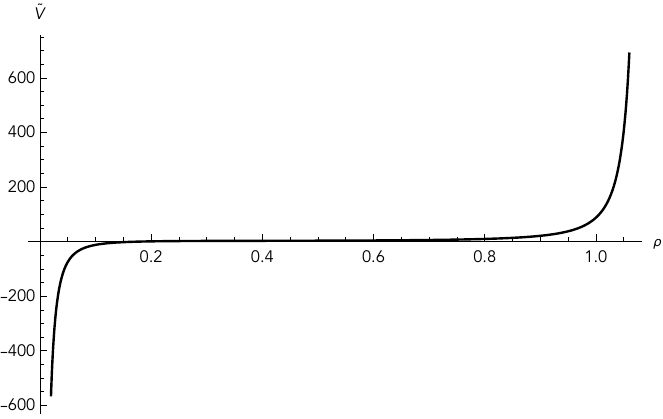} \hspace{1cm}
    \includegraphics[scale=0.55]{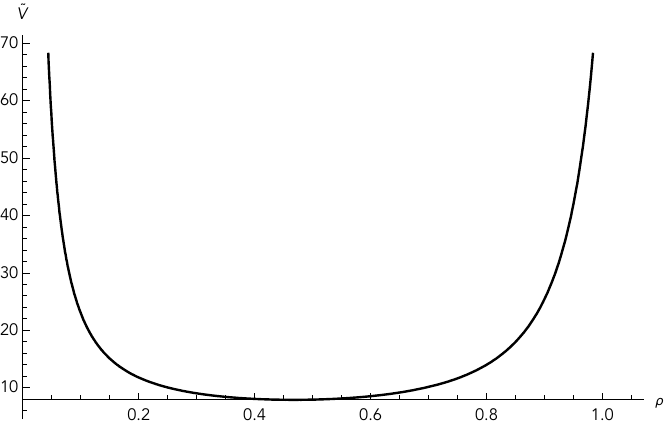}
    \caption{The effective potential $\tilde{V}(\r)$ plotted for $n=0$ (left) and $n=2$ (right) with $c=1, \mu=1$.}
    \label{fig:pot}
\end{figure}

We use the midpoint determinant method to numerically find the solutions of the eq.(\ref{schrof}) as described in \cite{Berg:2006xy}. We search for solutions that obey the  correct boundary conditions in the $r_1=r^*$ and $r_2\equiv r \to \infty$ boundaries. These solutions may exist only for specific values of $M^2$. 

For a second-order differential equation, the solution is characterised by its value and its derivative at a chosen point. One can form the vector $\left(\Psi_{i},\Psi'_{i}\right)$ in each boundary and evolve them numerically in the $r$ direction, hence they can be evaluated at other positions. Here $\Psi_{1}(r)$ will be the solution obtained from the boundary conditions in the boundary at $r_1$, etc. If for a particular value of $M^2$ one can find a solution that interpolates between the desired boundary conditions in the $r_1$ and $r_2$, this value of $M^2$ will be an eigenvalue of the equation. 
To perform the analysis, we form a matrix by putting the solution vectors $\left(\Psi_{i},\Psi'_{i}\right)$, coming from the $r_1$ and $r_2$ boundaries, next to each other. A certain value of $M^2$ will be an eigenvalue if and only if the determinant of this matrix, evaluated at some meeting point, is zero.

The spectra of fluctuations for different values of $n, c$ and $\m$ are plotted in Figure \ref{fig:spec}. The fluctuations have positive mass, indicating that there is no tachyonic instability in the background as long as these specific spin-two fluctuation modes are considered.

\begin{figure}[h]
    \centering
    \includegraphics[scale=0.5]{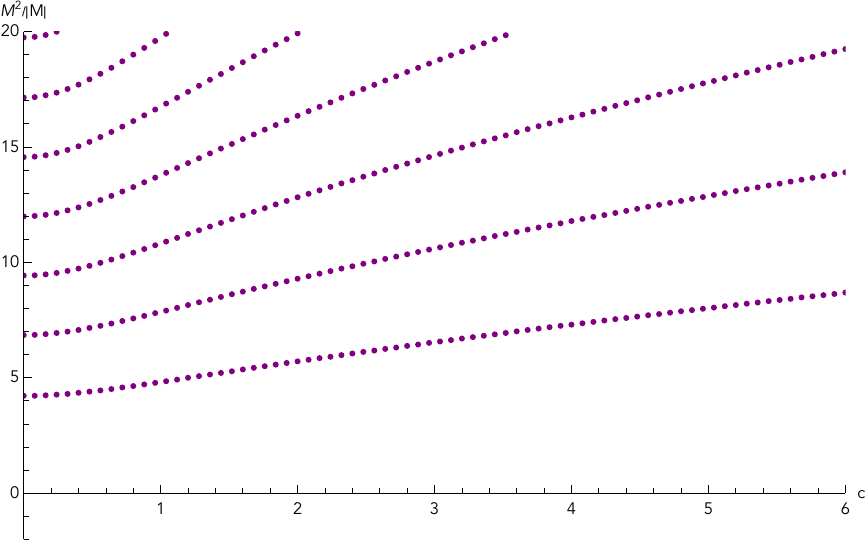} \hspace{1cm}
    \includegraphics[scale=0.55]{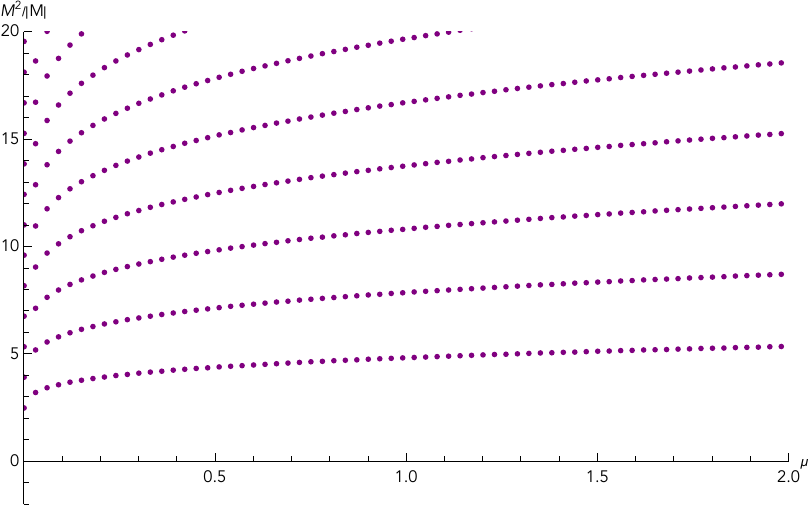}
    \includegraphics[scale=0.55]{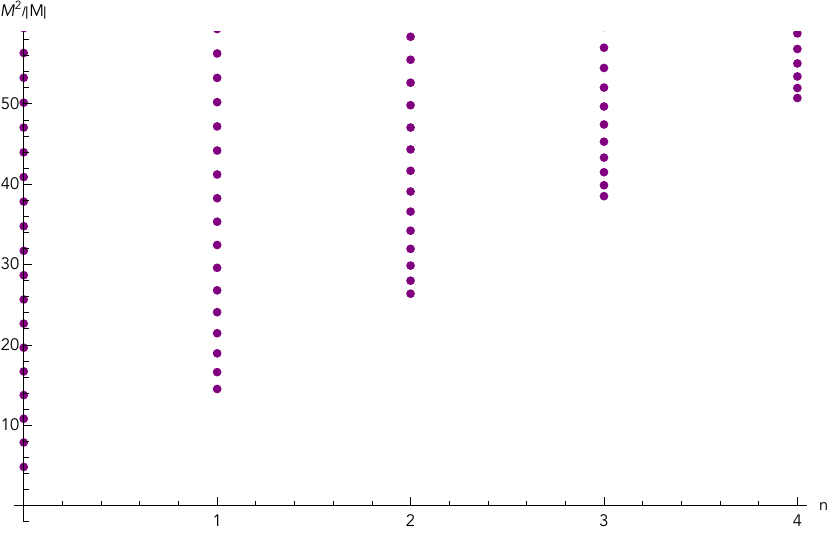}
    \caption{The spectrum of spin-two fluctuations. Top Left: Variation of  $c$ parameter with $\mu=1$. Top Right: Variation of  $\mu$ parameter with $c=1$. Bottom: Variation of n with $c=1, \mu=1$.}
    \label{fig:spec}
\end{figure}
\subsubsection{A stable background?}
Whilst the spectrum of masses for spin-two glueballs considered in the previous section gives positive values for $M^2$, this is hardly a proof of the stability of our backgrounds defined in eqs.(\ref{tendconfig})-(\ref{bc}).

In this paper, we do not study the stability of our backgrounds, but we will write some comments in this section.

A complete fluctuation analysis of our system is a very daunting task involving fluctuating all the Ramond and NS fields, and also the action of the localised flavour branes. It may be a good idea to restrict the number of fluctuating modes. This can be done by studying a reduction of 10d background to 6d-$SU(2)$ gauged supergravity  as explained in Appendix \ref{appendix1}. We can study fluctuations in six dimensions, keeping "less" fields.

In this six-dimensional context, we can attempt to write first-order equations (solving the second-order counterparts). Even when these are not the usual BPS equations (associated with SUSY). If this succeeds, one can attempt a "fake-supergravity" type argument for stability; see \cite{Freedman:2003ax} and references therein.

We wish to close with an argument used in a similar context \cite{Aharony:2002vp}.
Our QFT is a gapped system. Indeed, as argued in Section \ref{sec:holcent}, the quantity $c_{flow}$, vanishes in the far IR; hence, there is a mass gap--see below eq.(\ref{cholflow4}). Our deformation of the SUSY background (the solution for $r \to \infty$), is not small. At best, one can think that the solution is stable for large values of r as it is nearly BPS. 
This argument can not be continued to the $r\sim r^*$ region, as the deformation from CFT$_5$ dual background is large. On the other hand, at smaller values of $r$ as the dual QFT is gapped, one may expect that a small fluctuation in the background will not have a negative mass.

In summary, the study of fluctuations over our solution is a theme of interest that we plan to work on in the future.

\section{Conclusions}\label{conclx}
Let us begin with a brief summary of the contents of this paper.

We started in Section \ref{sec:background} presenting a new family of non-SUSY supergravity backgrounds in Type IIB. The backgrounds asymptote to AdS$_6\times S^2\times \Sigma_2$ for large values of the radial coordinate and a smooth manifold for small values of $r$. The family of backgrounds is smooth, except close to the location of the branes. The Page charges indicate an array of D5-NS5-D7 branes, typical of Hanany-witten set ups associated with five dimensional ${\cal N}=2$ field theories.

In Section \ref{sec:QFT} we proposed  a family of field theories dual to the family of backgrounds in Section \ref{sec:background}. These are strongly coupled ${\cal N}=1$ (eight Poincare SUSYs) five dimensional SCFTs that spontaneously compactify on a circle giving VEvs to a symmetry current and to the energy momentum tensor. We indicate a way to calculate the number of states in the QFT, obtaining a monotonic quantity that is vanishing in the IR (suggesting a gapped system) and asymptotes to the free energy of the five dimensional SCFT at high energies.

Wilson loops and Entanglement Entropy give indications of a behaviour interpolating between conformal-confining-screening.

We gave a proposal to calculate the Holographic Complexity of our system. This proposal shows the usual divergences, together with the contribution of the 5d SCFT and the influence of the IR gap-scale.

Masses of spin-two fluctuations on the Minkowski$_4$ directions are computed. They all have positive mass, hence do not destabilise the background.

In the future, it would be interesting to learn more about:
\begin{itemize}
    \item{The transition between conformal-confining-screening. In particular study Wilson loops in more generic configurations, as suggested above $r(x)$ and $\eta(x)$ should be studied. Similarly, the Entanglement Entropy should be studied with an eight surface that shows dependence $r(x_1,\eta,\sigma)$. These lead to interesting numerical problems. }
    \item{It should be interesting to study all other small fluctuations of the background, or at least study them in the reduced 6d gauged supergravity system. Ascertaining the positivity of the masses of these fluctuations is important. Otherwise, to prove the positivity of any fluctuation.}
    \item{It is interesting to find a solution like ours (with an IR end of the space),  preserving some amount of SUSY. This would put us in a more direct relation with the Anabal\'on-Ross systems. Some of the formalisms developed could be used.}
    \item{It would be interesting to use the procedure in this paper to find holographic duals to deformations of SCFTs in different dimensions.}
    \item{In Section \ref{flowcentral}, we found a monotonic c-function that interpolates between the gapped four dimensional IR and the conformal and SUSY UV five dimensional fixed point. It is of interest to understand how other recently proposed c-functions \cite{Caceres:2023mqz} work in this case. There may be some unity in these different definitions.  }
    \item{By Wick rotating our solutions, as explained in Appendix \ref{6dromans} one finds  Black branes after lifting to Type IIB. The study of the properties of these black objects is of interest.}
    \item{Our system contains the backreaction of localised D7 flavour branes. The calculation of meson masses in the present family of backgrounds is of interest. Specifically, in the line of the works \cite{Elander:2022ebt,Elander:2023aow}, it is interesting to search for light mesonic states being candidates for the Higgs particle in the framework of the Composite Higgs models.}
    \item{It would be interesting to understand from the viewpoint of the five dimensional brane web the deformation by VEVs in eqs.(\ref{VEVs1})-(\ref{VEVs2}). Similarly, if Matrix Model methods could calculate some observables along the RG flow, this would be very interesting.}
\end{itemize}
We hope to report on some of these in the near future.

\section*{Acknowledgments} For discussions, comments on the manuscript and for sharing their interesting ideas with us, we wish to thank: Andr\'es Anabal\'on, Elena Caceres, Andrew Frey, Yolanda Lozano, Simon Ross, Leonardo Santilli, Jiayue Yang.
The work of A.F. is supported by the STFC Consolidated Grant No. ST/V507143/1 and EPSRC Standard Research Studentship (DTP)
EP/T517987/1.
The work of C. N. is supported by grants ST/X000648/1 and ST/T000813/1.

{\bf Research Data Access Statement}---The data generated for this manuscript can be downloaded from
Ref.~\cite{Data}. 

{\bf Open Access Statement}---For the purpose of open access, the authors have applied a Creative Commons 
Attribution (CC BY) licence  to any Author Accepted Manuscript version arising.

\appendix

\section{The Supergravity Background Origin}\label{appendix1}
In this appendix, we describe the procedure to obtain the background in eq.(\ref{tendconfig}). These solutions are found in 6d Romans gauged supergravity  \cite{Romans:1985tw}, then uplifted to Type-IIB solutions. Below, we explain the details of this procedure.
\subsection{Six dimensional Romans $F_4$ supergravity}\label{summaryF4}
Here we review the content of six dimensional Romans'  $F_4$ gauged supergravity  \cite{Romans:1985tw}. The bosonic content of this supergravity, besides the vielbein, is a real scalar field $X$, a non-Abelian SU(2) gauge field $A^i$ with field strength 
\begin{equation}
F^i= dA^i + \frac12 \epsilon^{ijk} A^j \wedge A^k,
\end{equation} 
a three-form 
\begin{equation}
F_3= dA_2 \, ,\label{eq1}
\end{equation} 
and an Abelian gauge field $A_1$ with curvature 
\begin{equation}
F_2= dA_1+ \frac{2}{3} \tilde{g} A_2 \, .\label{eq2}
\end{equation} 
$\tilde{g}$ is a coupling parameter in the 6d theory. 

We write the bosonic part of the Lagrangian,
\begin{eqnarray}
& & \mathcal{L} = R *_6 1+ 4 \frac{*_6 d X \wedge d X}{X^2}- \tilde{g}^2 \left(\frac29 X^{-6} - \frac83 X^{-2}-2 X^2\right) *_6 \label{eq:lagrangian6}\\
& & \qquad  + \frac12 X^4 *_6 F_3 \wedge F_3  - \frac12 X^{-2} \left(*_6 F_2 \wedge F_2 + \frac1{\tilde{g}^2} *_6 F^i \wedge F^i\right) \nonumber\\
& & \qquad -\frac12 \tilde{A}_2 \wedge \left( d A_1 \wedge d A_1 + \frac23 \tilde{g} d A_1 \wedge A_2 + \frac{4}{27} \tilde{g}^2 A_2 \wedge A_2 + \frac1{\tilde{g}^2} F^i \wedge F^i \right). \nonumber
\end{eqnarray}
The equations of motion read
\begin{eqnarray}
& & d(X^4 *_6 F_3) = \frac12 F_2 \wedge F_2 + \frac{1}{2 \gt^2} F^i \wedge F^i + \frac23 \gt X^{-2} *_6 F_2, \label{eqs6d} \\
& & d(X^{-2} *_6 F_2) = - F_2 \wedge F_3 \\
& & D(X^{-2} *_6 F^i)= - F_3 \wedge F^i \\
& & d(X^{-1} *_6 d X) = \frac14 X^4 *_6 F_3 \wedge F_3  - \frac{X^{-2}}{8} \left(*_6 F_2 \wedge F_2 + \frac1{\tilde{g}^2} *_6 F^i \wedge F^i\right) \nonumber \\[2mm]
& & \qquad \qquad \qquad \quad  - \tilde{g}^2 \left(\frac16 X^{-6}  - \frac23 X^{-2}+ \frac12 X^2\right) *_6 1. \label{eqs6dlast}
\end{eqnarray}
Here, $D$ is defined as the SU(2) gauge-covariant derivative. Its action on a differential form $C^i$ is defined by
\begin{equation}
DC^i= dC^i +\epsilon_{ijk}A^j \wedge C^k \, .\label{eq3}
\end{equation}
The Einstein's equations are
\begin{eqnarray}
& & R_{\mu \nu} = 4 X^{-2} \partial_\mu X \partial_\nu X+ \gt^2   \left(\frac{1}{18} X^{-6}  - \frac23 X^{-2}- \frac12 X^2\right) g_{\mu \nu} + \frac{X^4}{4} \left(F_{3 \, \mu} \cdot F_{3 \, \nu} - \frac{1}{6} g_{\mu \nu} F_3^2\right) \nonumber \\[2mm]
& & \qquad + \frac{X^{-2}}{2} \left(F_{2 \, \mu} \cdot F_{2 \, \nu} - \frac{1}{8} g_{\mu \nu} F_2^2\right) + \frac{X^{-2}}{2 \gt^2} \left(F^i_{2 \, \mu} \cdot F^i_{2 \, \nu} - \frac{1}{8} g_{\mu \nu} (F_2^i)^2\right), \label{Einstein6d}
\end{eqnarray}
With $F_\mu = \iota_\mu F$ being the contraction with the vector $\partial_\mu$, $F \cdot G = F_{\mu_1 \dots \mu_p} G^{\mu_1 \dots \mu_p}$, and $F^2 = F \cdot F$.

The fermionic part of the Lagrangian is written in \cite{Romans:1985tw}.
Setting all matter fields to zero $A_2=A_1=A^i=0$ and $X=1$ leads to a simple solution. The metric is  AdS$_6$ with radius $R^2=\frac{2}{9}\gt^2$ and the solution preserves eight Poincar\'e supercharges.

Let us  now discuss a specific solution to this six dimensional system.

\subsection{The background in 6d Romans Supergravity}\label{6dromans}
We study a background which has a nontrivial background metric, supported by only a single component of the SU(2) Yang-Mills fields $F_2^{(3)}$ and a dilaton \cite{Cvetic:1999un},
$$
\begin{aligned}
ds_6^2 &=& -H(r)^{-3/2}\, f(r)\, dt^2 + H^{1/2}\, (f(r)^{-1}\, dr^2 + r^2\,
d{\vec x}^2_{4}) \,,\nn\\
\phi&=& \ft 1{\sqrt2}\, \log H(r)\,,\quad A^{3}_1 = \sqrt{2} (1-H(r)^{-1})
\,  \frac{\sqrt{\mu}}{c}\, dt\,,\nn\\
f(r) &=& -\fft{\mu}{r^3} + \ft29 g^2\, r^2\, H(r)^2\,,\qquad H(r)=1 + \fft{c^2}{\, r^3}\,,\nn
\end{aligned}
$$
where the dilaton $\phi$  is related to the $X$ field by
$X=e^{-\fft1{2\sqrt2}\phi}$. There are two free parameters $\mu$ and $c$ appearing in the solution.

\subsection{Double Wick rotation}
After performing a double wick rotation on internal coordinates as $t\rightarrow i\phi$, $\phi \rightarrow i t$ and analytic continuation of the $c$ parameter we have our new background:
\begin{align} \label{6dBackground}
ds_6^2 &= H^{1/2}\, (f(r)^{-1}\, dr^2 + r^2\,(
-dt^2+dx_1^2+dx_2^2+dx_3^2))+ H(r)^{-3/2}\, f(r)\, d\phi^2  \,,\\
\phi&= \ft 1{\sqrt2}\, \log H(r)\,,\quad A^{3}_1 = \sqrt{2} (1-H(r)^{-1})
\,  \frac{\sqrt{\mu}}{c}\, d\phi\,,\nn\\
f(r) &= -\fft{\mu}{r^3} + \ft29 g^2\, r^2\, H(r)^2\,,\qquad H(r)=1 - \fft{c^2}{\, r^3}\,,\nn
\end{align}

By compactifying the $\phi$ coordinate, as the result of the term $H(r)^{-3/2}\, f(r)\, d\phi^2$ in the metric, the $\phi$ coordinate can shrink to zero size at the roots of $f(r)$. One needs to consider $r^*$, the largest positive root of $f(r)$ and pick an appropriate period for this cycle to have a smoothly closing geometry in the $(r,\phi)$ plane which is done in eq.(\ref{Period}).

\subsection{SUSY variations}\label{appendix2}
Here, we study the supersymmetric properties of our background. The fermionic content of the Romans gauged supergravity are four gravitini $\psi_{\m\, i} $ and four spin 1/2 fields $\c_i$. The SUSY variations for these fermionic degrees of freedom with only $SU(2)$ gauge field and the dilaton turned on, is
\ba
\d\c_i &=& \left( \fr1{\sqrt{2}}\, \g^\m \, 
\pa_\m\f + A \, \g_7 \right) \e_i + \fr1{2\sqrt{2}}\,\g^{\m\n}\,
({\hat{H}}_{\m\n})_i^{\,\,\, j} \, \e_j \,\,\, ,\label{gautransf}\\
\d\psi_{\m\, i} & = & \left( \nabla_\m  + T \, \g_\m \, \g_7 \right) \, \e_i   + \left( g \, A_\mu^I \, (T^I)_i^{\,\,\, j}
-\frac{1}{4\sqrt{2}} \, (\gamma_\mu^{\,\,\,\, \nu\rho} - 6 \, \delta_\mu^{\,\,\,\, \nu} \, \gamma^\rho)
({\hat{H}}_{\nu\rho})_i^{\,\,\, j} \right) \epsilon_j \,\,\, ,
\label{gratransf}
\ea
where $A$, $T$, and $\hat{H}$ are defined as
\ba
A &\equiv& \frac{1}{4\sqrt{2}} \, ( g \, \rme^{\frac{\phi}{\sqrt{2}}}
- 
g \, \rme^{\frac{-3\phi}{\sqrt{2}}}),\;\;\;\;\;\;\;\;\;\; 
T \equiv - \frac{1}{8\sqrt{2}} \, (g \,
e^{\frac{\phi}{\sqrt{2}}} + g/3 \, \rme^{\frac{-3\phi}{\sqrt{2}}}) \,\,\,
,\label{atdef}\\
({\hat{H}}_{\mu\nu})_i^{\,\,\, j} &\equiv&
\rme^{-\frac{\phi}{\sqrt{2}}} \, \left( \gamma_7 \,
F_{\mu\nu}^I \, (T^I)_i^{\,\,\, j} \right) \,\,\,.
\label{hdef}
\ea
The gauge-covariant derivative ${\cal D}_\m$ acting on the Killing spinor is 
\beq
{\cal D}_\m\,\e_i = \nabla_\m\,\e_i + g\,A^I_\m\,(T^I)_i^{\;\;j}\,\e_j \,\,\, ,
\eeq
with
\beq
\nabla_\mu \e_i \equiv (\partial_\mu+\frac{1}{4} \, 
\omega^{\, \, \, \, \alpha \beta}_\mu  \,
      \gamma_{\alpha \beta} ) \, \e_i \,\,\, ,
\eeq
Indices $\a, \b$ are tangent space (flat) indices,
while $\m, \n$ are space-time indices. By inspection, we realize that our solution in the previous section does not preserve any SUSY unless $\mu=0$ and $c=0$ which leads us to the pure $AdS_6$ background.

\subsection{Uplift to Type IIB}
It is shown in \cite{Legramandi:2021uds}, \cite{Hong:2018amk} that solutions to 6d Romans supergravity possess an uplift to an infinite family of solutions in Type IIB supergravity. Our solution with nontrivial metric, dilaton and SU(2) field lifts to a configuration in type IIB  given by,
\begin{eqnarray}
& & d s^2_{st} = f_1 \left( d s^2_6 + f_2 d s^2 (\tilde{S}^2)+ f_3 d s^2 (R^2) \right) \, \label{tendconfig1}\\[2mm]
& &C_0 = f_7,\;\;\;\;  e^{-2\Phi} = f_6, \qquad F_5 = 4 (G_5 +*_{10} G_5), \nonumber\\[2mm]
& &B_2 = f_4\text{Vol}(\tilde{S}^2) - \frac{2}{9} \eta y^i F^i \, , \nonumber\\[2mm]
& & C_2 = f_5\text{Vol}(\tilde{S}^2)-4  \partial_\sigma (\sigma V) y^i F^i \, . \nonumber
\end{eqnarray}
Here $d s^2_6$ is defined in terms of the 6d gauged supergravity background metric as 
\begin{equation}
d s^2_6=  \frac{2\tilde{g}^2}{9}  d s^2_{\text{gauged sugra.}}
\end{equation}
The functions $f_i$ read:
\begin{eqnarray}
& & f_1 =\frac{2}{3 X^2} \left( \sigma^2 + \frac{3 X^4\sigma \partial_{\sigma}V}{\partial_{\eta}^2 V}\right)^{1/2}, \qquad f_2 = \frac{X^2\partial_{\sigma}V\partial_{\eta}^2V}{3{\Lambda}}, \qquad  f_3 = \frac{X^2\partial_{\eta}^2V}{3\sigma\partial_{\sigma}V} \label{defi2} \\
& &   f_6= (18)^2 \frac{{3 X^4 (\sigma^2 \partial_\sigma V) (\partial_{\eta}^2 V) }}{\left(3 X^4 \partial_\sigma V+\sigma  \partial_{\eta}^2 V\right)^2} \Lambda , \qquad f_7= 18 \left(\partial_\eta V+\frac{3 X^4 \sigma  \partial_\sigma V \partial_{\sigma \eta}^2V}{3 X^4 \partial_\sigma V+\sigma  \partial_{\eta}^2 V}\right), \nonumber\\
& & f_4= \frac{2}{9} \left(\eta -\frac{\sigma  \partial_{\sigma} V \partial_{\sigma \eta}^2 V}{\Lambda }\right),\qquad f_5=4 \left(V-\frac{\sigma   \partial_{\sigma} V \left( \partial_{\eta} V \partial_{\sigma \eta}^2 V-3 X^4 \partial_{\eta}^2 V \partial_{\sigma} V \right)}{\Lambda}\right),\nonumber\\
& & \Lambda=3 X^4 \partial _{\eta}^2 V \partial _{\sigma }V+ \sigma \left[\left(\partial^2_{\eta \sigma} V \right)^2+\left(\partial _{\eta}^2 V\right)^2\right] .\nonumber
\end{eqnarray}
$\tilde{S}^2$  sphere is fibered over the 6d spacetime, 
\begin{equation}
\text{Vol}(\tilde{S}^2) =\epsilon^{ijk} y^iDy^j\wedge Dy^k , \qquad d s^2_{\tilde{S}^2}= Dy^i Dy^i ,\label{eq4}
\end{equation}
where $y^i$ are the embedding coordinates of the sphere $S^2$ and can be chosen as
\begin{equation}
\label{eq:S2_embedding}
y^1= \sin\theta\sin\varphi_1, \qquad y^2= \sin\theta \cos\varphi_1, \qquad y^3= -\cos\theta . \\
\end{equation}
The symbol $D$ represents the covariant derivative
\begin{equation}
DC^i= dC^i +\epsilon_{ijk}A^j \wedge C^k \, .\label{eq3bis}
\end{equation}
$G_5$ is a differential form:
\begin{equation}
\label{eq:G5_def}
G_5 = -\frac{2}{3 X^2} (*_6 F^i) \wedge D (y^i \sigma ^2 \partial_\sigma V) \, .
\end{equation}
The 10d metric after simplifying the eq.(\ref{eq4}) reads
\begin{eqnarray}
& & d s^2_{st} = f_1 \left( d s^2_6 + f_2 (d \theta^2 +\sin{\theta}^2(d \varphi_1-A)^2)+ f_3 (d \sigma^2+d \eta^2) \right) \, 
\end{eqnarray}
Following the procedure, the reader can check that the full configuration in eq.(\ref{tendconfig}) arises.

\subsection{Uplift to massive Type IIA}
In the body of the paper, we have worked with the Type IIB lift of the solution in eq.(\ref{6dBackground}). We can also lift this solution to massive Type IIA theory. The background for our solution reads \cite{Cvetic:1999un}
\begin{align}
d\hat s_{10}^2 =& (\sin\xi)^{\fft1{12}}\, X^{\fft18}\Big[
\Delta^{\fft38}\, ds_6^2 + 2g^{-2}\, \Delta^{\fft38}\, X^2\, d\xi^2
+\ft12g^{-2}\, \Delta^{-\fft58}\, X^{-1}\, \cos^2\xi
\sum_{i=1}^3(\sigma^i - g\, A^i)^2\Big]\,,\nn\\
\hat F_\4 =& -\ft{\sqrt2}{6}\, g^{-3}\, s^{1/3}\, c^3\, \Delta^{-2}\,
U\, d\xi\wedge\ep_\3 -\sqrt2 g^{-3}\, s^{4/3}\, c^4\, \Delta^{-2}\,
X^{-3}\, dX\wedge \ep_\3 \nn\\
& +\ft1{\sqrt2} g^{-2}\,
s^{1/3}\, c\, F_2^{(3)}\wedge \, h^3\wedge d\xi -\ft1{2\sqrt2} g^{-2}\,
s^{4/3}\, c^2\, \Delta^{-1}\, X^{-3}\,  F_2^{(3)} \wedge
h^1\wedge h^2\,\\
e^{\hat\phi} =&  s^{-5/6}\, \Delta^{1/4}\, X^{-5/4} \qquad F_0=m=\ft{\sqrt2}{3}\, g\,,\nn   
\end{align}
where
\bea
\Delta &\equiv & X\cos^2\xi +X^{-3} \sin^2 \xi\,,\nn\\
U &\equiv& X^{-6}\, s^2 - 3 X^2\, c^2 + 4 X^{-2}\, c^2 - 6 X^{-2}\,.
\eea 
$\sigma^i$ are left-invariant 1-forms on $S^3$, 
satisfying $d\sigma^i = -\ft12 \ep_{ijk}\, \sigma^j\wedge \sigma^k$ and $h^i\equiv \sigma^i-g\, A_\1^i$, $\ep_\3\equiv
h^1\wedge h^2\wedge h^3$.  We have used the shorthand notation $s=\sin\xi$ and $c=\cos\xi$.  
The mass parameter $m$ of the
massive type IIA theory is related to the gauge coupling $g$ by $m= \ft{\sqrt2}{3}\, g$.

The reason the Type IIB lift is chosen in the body of the paper, is the ability to work with an infinite family of backgrounds- one member of the family for each function $V(\sigma,\eta)$. Besides, the dual field theory picture of the lift in massive IIA (involving O8 planes and an exceptional group E$_{N_f+1}$ global symmetry) is less clear than the one presented in the body of the paper, based on unitary gauge and flavour groups.

Finally, we worked with the Type IIB lift because the massive IIA background contains an overall warp factor that vanishes at $\xi =0$, with a divergent dilaton at the same point. The supergravity is not a good approximation close to $\xi=0$ and the calculations exploring that region in the context of massive Type IIA are not trustable.

\section{Near Boundary Expansions} \label{appendix:Boundary}
In this appendix, we analyse the expansion of our background metric and fields near the $r\to \infty$ boundary to study the deformations at the UV fixed point of the dual field theory. We perform this in the 6d supergravity system discussed in Appendix \ref{appendix1}. The QFT data that we need are the operators turned on by these deformations, and their vacuum expectation value. Since the boundary metric 
couples to the boundary stress-energy tensor, the vacuum expectation values of this tensor can be obtained from the asymptotics of the background metric and other coupled scalars. Following \cite{ Skenderis:2002wp, deHaro:2000vlm}, the bulk metric near the boundary of an asymptotically AdS background takes the form,
\be \label{grahamff}
ds^2 = {1 \over r^2} (dr^2 + g_{ij}(x,r) dx^i dx^j)
\ee
in some appropriately chosen coordinate system with radial direction r. These are the so-called Fefferman-Graham coordinates and the boundary is located at $r \to 0$. The tensor
$g_{ij}(x,r)$ can be written as
\be
g_{ij}(x,r) = g_{(0)ij} + r g_{(1) ij} + r^2 g_{(2)ij} + ...
\ee
One can find the coefficients $g_{(k) ij}, k>0$
from Einstein's equations. 
 We introduce the new coordinate
$\r=r^2$.  In these coordinates we have
\bea \label{coord}
&ds^2=G_{\m \n} dx^\m dx^\n = {d\r^2 \over 4 \r^2} + 
{1 \over \r} g_{ij}(x,\r) dx^i dx^j,\\ \nonumber
& g(x,\r)=g_{(0)} + \cdots + \r^{d/2} g_{(d)} + h_{(d)} \r^{d/2} \ln \r + ... 
\eea

Near the boundary, different fields in the background have an asymptotic expansion of the form
\be \label{assympexpansion}
{\cal F}(x,\r) = \r^m \left(f_{(0)}(x) + f_{(2)}(x) \r + \cdots +
\r^n (f_{(2 n)}(x) + \log \r \tilde{f}_{(2 n)}(x)) + ... \right)
\ee
The equations of motion are second-order differential equations 
in $\r$, hence two independent solutions can be found. The asymptotic behaviours for these solutions are chosen to be $\rho^{m}$ and $\rho^{m+n}$.

The boundary field $f_{(0)}$  multiplying the leading term, 
$\r^m$, is interpreted as the source for the dual operator inserted in the field theory. The coefficient $f_{(2 n)}$ is associated with the 1-point function of the corresponding operator. 

In our case, the boundary is five-dimensional. Following \cite{deHaro:2000vlm}, the expectation value for the boundary stress-energy tensor, after performing holographic renormalisation, reads
\be \label{Tij}
\langle T_{ij} \rangle= 
5l^4 g_{(5)ij}
\ee
where $l$ is AdS radius related to
the cosmological constant as $\Lambda= -\frac{10}{l^2}$ and we set $16\pi G_N\equiv1$.

The metric in eq.(\ref{6dBackground}) is not in the required form of eq.(\ref{grahamff}). We perform some coordinate transformations. First we take $r\to 1/z$ which brings the boundary from $r \to \infty$ to $z \to 0$,
\begin{align} \label{6dBackgroundz}
ds_6^2 &= H(z)^{1/2}\, \left(\frac{1}{f(z)z^4}\, dz^2 + \frac{1}{z^2}\,(
-dt^2+dx_1^2+dx_2^2+dx_3^2)\right)+ H(z)^{-3/2}\, f(z)\, d\phi^2  \,.
\end{align}
Now we we expand the metric components near $z \to 0$ and perform a coordinate change $z \to \bar r (z)$  to bring the metric to the Fefferman-Graham form, at least asymptotically up to the required order. The coordinate change with its inverse is (setting $\gt=\sqrt{\frac{9}{2}}$ for simplicity)
\beqs
\bar r(z) &= z+\frac{c^2 z^4}{4}+\frac{\mu  z^6}{10}+O\left(z^7\right),\\
z(\bar r) &= \bar r-\frac{c^2 {\bar r}^4}{4}+\frac{\mu  {\bar r}^6}{10}+O\left({\bar r}^7\right).
\eeqs
The expansion of time and space components of the metric in this coordinate system will be
\beqs
g_{tt}(\bar r) & = -\frac{1}{{\bar r}^2}-\frac{\mu  {\bar r} ^3}{5}+O\left({\bar r}  ^4\right),\\
g_{x_ix_i}(\bar r) & = \frac{1}{{\bar r}^2}+\frac{\mu  {\bar r} ^3}{5}+O\left({\bar r}  ^4\right),\\
g_{\phi\phi}(\bar r) & = \frac{1}{{\bar r}^2}-\frac{4\mu  {\bar r} ^3}{5}+O\left({\bar r} ^4\right).
\eeqs
Hence, one can read the VEVs for boundary stress-energy tensor using eq.(\ref{Tij}),
\begin{equation} \label{boundaryT}
    \langle T_{tt} \rangle = -\mu , \quad  \langle T_{x_i x_i} \rangle = \mu , \quad \langle T_{\phi\phi} \rangle = 4\mu 
\end{equation}
One can also find an asymptotic expansion for the $X$ and $A_1^{(3)}$ in this background
\begin{align}
& X(\bar r)-1  = c_2 {\bar r}^2 + c_3 {\bar r}^3 - 11 c_2^2/2 {\bar r}^4 + 1/2 (c^2 c_2 - 6 c_2 c_3) {\bar r}^5+ O\left({\bar r}  ^6\right), \\
& A_1^{(3)}(\bar r)  = a_0 + a_3 {\bar r} ^3 + O\left({\bar r}  ^6\right).
\end{align}
By comparing to our solution's expansion in eq.(\ref{6dBackground}) 
\beqs
X(\bar r)-1 & = \frac{c^2}{4} {\bar r}^3 + O\left({\bar r}  ^6\right), \\
A_1^{(3)}(\bar r) & = -3c\sqrt{\mu} {\bar r} ^3 + O\left({\bar r}  ^6\right),
\eeqs
we realize that the subleading modes in the expansions are turned on, hence
\begin{equation} \label{boundaryXA}
    \langle J \rangle = -3c\sqrt{\mu}  , \quad  \langle {\cal O}_X \rangle = \frac{c^2}{4}.
\end{equation}
$X$  leads to the insertion of an operator of dimension three operator with the quoted VEV and $A_1^{(3)}$ couples to a background global R-symmetry current in the boundary which can be interpreted as a background Wilson loop insertion in the QFT.

\section{Some useful identities}\label{usefulidentities}
In this appendix we mention some useful identities mostly derived in \cite{Legramandi:2021uds} and extend them. We consider the potential defined by eq.(\ref{eq:fourier_vhat}) and the relations,
\begin{eqnarray}
& & \hat{V}= \sum_{k=1}^\infty  a_k \sin\left(\frac{k \pi\eta}{P}\right) e^{-\frac{k \pi |\sigma|}{P}},\label{identities1}\\
& & \partial_\sigma \hat{V}=-\sum_{k=1}^\infty  a_k\left(\frac{k \pi}{P}\right) sg(\sigma)\sin\left(\frac{k \pi\eta}{P}\right) e^{-\frac{k \pi |\sigma|}{P}},\nonumber\\
& & \partial_\eta \hat{V}=\sum_{k=1}^\infty  a_k\left(\frac{k \pi}{P}\right) \cos\left(\frac{k \pi\eta}{P}\right) e^{-\frac{k \pi |\sigma|}{P}},\nonumber\\
& & \partial_\sigma\partial_\eta \hat{V}= 
-\sum_{k=1}^\infty  a_k\left(\frac{k^2 \pi^2}{P^2} \right)sg(\sigma)\cos\left(\frac{k \pi\eta}{P}\right) e^{-\frac{k \pi |\sigma|}{P}},\nonumber\\
& & \partial_\eta^2 \hat{V}= -\sum_{k=1}^\infty  a_k\left(\frac{k^2 \pi^2}{P^2} \right) \sin\left(\frac{k \pi\eta}{P}\right) e^{-\frac{k \pi |\sigma|}{P}},\nonumber\\
& & \partial_\sigma^2 \hat{V}= \sum_{k=1}^\infty  a_k\left(\frac{k \pi}{P} \right)\left(\frac{k\pi}{P} -\delta(\sigma) \right) \sin\left(\frac{k \pi\eta}{P}\right) e^{-\frac{k \pi |\sigma|}{P}}.\nonumber
\end{eqnarray}
Also,
\begin{eqnarray}
& & \partial_\sigma V= \frac{\sigma \partial_\sigma \hat{V} -\hat{V}}{\sigma^2},\;\;\;\;\partial_\eta V=\frac{\partial_\eta \hat{V}}{\sigma},\;\;\;\partial_\eta^2 V= \frac{\partial_\eta^2 \hat{V}}{\sigma}\label{identities2}\\
& & \partial_\sigma\partial_\eta V=\frac{\sigma \partial_\eta\partial_\sigma \hat{V} -\partial_\eta \hat{V}}{\sigma^2},\;\;\; \partial_\sigma^2 V=\frac{2\hat{V} -2 \sigma \partial_\sigma \hat{V} +\sigma^2\partial_\sigma^2\hat{V}}{\sigma^3}.\nonumber
\end{eqnarray}
These results are used extensively in the study of the behaviour of the potentials and field strengths as needed in the Page charges calculations in Section \ref{pagesection}.
\subsection{The field $B_2$}
We will use the identities above (\ref{identities1})-(\ref{identities2}), to study the $B_2$ field in $r \to \infty$ limit. In this limit $X(r) \sim 1$ and  hence the expressions for functions $f_i$ simplify. Ignoring the volume of the two-sphere $\text{Vol}(S^2)=\sin\theta d\theta \wedge d\phi$, the expression is,
\begin{eqnarray}
& & \frac{2}{\pi} B_2= \eta -\frac{(\sigma \partial_\sigma V)( \partial_\sigma\partial_\eta V)}{\Lambda}=\eta-\frac{( \sigma^2 \partial_\sigma \hat{V} -\sigma \hat{V})( \sigma\partial_\sigma\partial_\eta \hat{V} -\partial_\eta \hat{V})}{\Lambda \sigma^4}.\nonumber\\
& & \sigma^4 \Lambda= \sigma \left[ (\sigma\partial_\sigma\partial_\eta \hat{V} -\partial_\eta^2\hat{V})^2 + (\partial_\eta^2\hat{V})( 3\sigma\partial_\sigma \hat{V} -3\hat{V} -\sigma^2\partial_\sigma^2\hat{V})\right]
\end{eqnarray}
Replacing the expansions in eqs.(\ref{identities1}) we find
\begin{eqnarray}
& & \sigma^4 \Lambda= \sigma \left( {\cal M}^2 + {\cal N} {\cal S}\right),\nonumber\\
& & {\cal M}=\sum_{k=1}^\infty a_k \left( \frac{k^2\pi^2}{P^2}\right)\left( \sin\left( \frac{k\pi \eta}{P}\right)  + |\sigma| \cos\left( \frac{k\pi \eta}{P}\right) \right) e^{-\frac{k \pi |\sigma|}{P}} ,\nonumber\\
& & {\cal N}= \sum_{k=1}^\infty a_k \left( \frac{k^2\pi^2}{P^2}\right)\sin\left( \frac{k\pi \eta}{P}\right) e^{-\frac{k \pi |\sigma|}{P}},\nonumber\\
& & {\cal S}=\sum_{k=1}^\infty a_k \sin\left( \frac{k\pi \eta}{P}\right) e^{-\frac{k \pi |\sigma|}{P}} \left( 3+\frac{3k\pi|\sigma|}{P} +\frac{k^2\pi^2\sigma^2}{P^2}\right) .\nonumber
\end{eqnarray}
For $\sigma\to\pm \infty$ one can check that we have 
\begin{equation}
\sigma \Lambda= a_1^2e^{-2\frac{\pi|\sigma|}{P}}\frac{\pi^4}{P^4}.\label{lambdaidentity}
\end{equation}
Hence, for $B_2$ the result is,
\begin{eqnarray}
& &\frac{2}{\pi}  B_2-\eta= \frac{{\cal P}{\cal Q}}{\Lambda \sigma^4},\label{B2identity}\\
& &  {\cal P}= \sum_{k=1}^\infty a_k \sin\left( \frac{k\pi \eta}{P}\right) \sigma e^{-\frac{k \pi |\sigma|}{P}} \left( 1+\frac{k\pi|\sigma|}{P}\right),\nonumber\\
& & {\cal Q}= \sum_{k=1}^\infty a_k \cos\left( \frac{k\pi \eta}{P}\right) \left(\frac{k\pi}{P}\right) e^{-\frac{k \pi |\sigma|}{P}} \left( 1+\frac{k\pi|\sigma|}{P}\right).\nonumber
\end{eqnarray}
Evaluating the field in the limit $\sigma\to\infty$ we have
\begin{equation}
B_2(\pm\infty,\eta)= f_4(\pm\infty,\eta)=\frac{\pi}{2} \left[\eta -\frac{P}{\pi} \sin\left( \frac{\pi \eta}{P}\right) \cos\left( \frac{\pi \eta}{P}\right) \right].\label{identityB2}
\end{equation}
In $r \to r^*$ case, $X(r) \neq 1$, so eq.(\ref{lambdaidentity}) will be modified to 
\begin{equation}
\sigma \Lambda= a_1^2e^{-2\frac{\pi|\sigma|}{P}}\frac{\pi^4}{P^4}(X(r^*)^4 \cos^2(\frac{\pi \eta}{P})+\sin^2(\frac{\pi \eta}{P})){P^4},\label{lambdaidentityp}
\end{equation}
which does not lead to a quantized NS5 charge --- see Section \ref{pagesection}.
\subsection{The field $C_0$}
In $r \to \infty$, $X(r)\sim1$, we have the expression,
\begin{eqnarray}
& & \frac{C_0}{2}=
\partial_\eta V+\frac{\sigma \partial_\sigma\partial_\eta V}{1+ \frac{\sigma \partial_\eta^2 V}{3\partial_\sigma V}}= \sum_{k=1}^\infty a_k \cos\left( \frac{k\pi \eta}{P}\right) e^{-\frac{k \pi |\sigma|}{P}}\left[\frac{k\pi}{P\sigma}  -\frac{\frac{k\pi}{P\sigma} +\frac{k^2\pi^2}{P^2}sgn(\sigma)    }{1+ \frac{{\cal C}}{{\cal B}}}\right],\label{C0exp}\\
& & {\cal C}=\sigma^2  \sum_{k=1}^\infty a_k \left( \frac{k^2\pi^2}{P^2}\right)\sin\left( \frac{k\pi \eta}{P}\right) e^{-\frac{k \pi |\sigma|}{P}},\;\;\;{\cal B}= 3 
\sum_{k=1}^\infty a_k \sin\left( \frac{k\pi \eta}{P}\right) e^{-\frac{k \pi |\sigma|}{P}}\left(1+\frac{k\pi |\sigma|}{P}\right).\nonumber  
\end{eqnarray}
One can check that at $\sigma=\epsilon$ for small $\epsilon$---and using
$2\pi k a_k= - P c_k$ one has
\begin{equation}
C_0(0,\eta)=f_7(0,\eta)= 9 \sum_{k=1}^\infty c_k \left( \frac{k\pi}{P}\right)\cos\left( \frac{k\pi \eta}{P}\right) = 9\partial_\eta {\cal R}(\eta).\label{identityC0}
\end{equation}
In the $r=r^*$ case, it can be shown in a similar manner that the same result is obtained and the $X(r)$ dependence drops in this calculation. 
\subsection{The combination $C_2- B_2 C_0$}
We  now investigate the expression appearing in the calculation of the Page charge for D5 branes in the limit $r \to \infty$. One can check that the same result applies to $r \to r^*$.
Ignoring the volume of the two-sphere  we have,
\begin{align}
\frac{C_2- B_2 C_0}{4}
=&
V- \eta\partial_\eta V +\frac{\sigma\partial_\sigma V -\eta\sigma \partial_\eta\partial_\sigma V}{1+\frac{\sigma \partial_\eta^2 V}{3\partial_\sigma V}}. \label{c2b2c01} \\[2mm]
\frac{\sigma(C_2- B_2 C_0)}{4}=&\hat{V}-\eta\partial_\eta \hat{V}   +\left[\frac{\sigma\partial_\sigma \hat {V} -\eta\sigma \partial_\eta\partial_\sigma \hat{V} -\hat{V} +\eta\partial_\eta \hat{V}}{1+\frac{\sigma^2 \partial_\eta^2 \hat{V}}{3\sigma \partial_\sigma \hat{V} - 3\hat{V} }}\right]\nonumber\\[2mm]
=&\sum_{k=1}^\infty a_k\left[ \sin\left( \frac{k\pi \eta}{P}\right)  -\left( \frac{k\pi\eta}{P}\right)\cos\left( \frac{k\pi \eta}{P}\right)  \right]e^{-\frac{k \pi |\sigma|}{P}}\left(1- \frac{1+ \frac{k \pi |\sigma|}{P}}{1+\frac{{\cal C}}{ {\cal B}} } \right). \nonumber
\end{align}
${\cal C},{\cal B}$ have been introduced in eq.(\ref{C0exp}).
We now have,
\begin{eqnarray}
& & (C_2- B_2 C_0)\Big]_{\sigma=\epsilon}^{\sigma=-\epsilon}= f_5- f_7 f_4\Big]_{\sigma=\epsilon}^{\sigma=-\epsilon}=8 \sum_{k=1}^\infty a_k \frac{k \pi}{P}\left[ \sin\left( \frac{k\pi \eta}{P}\right)  -\left( \frac{k\pi\eta}{P}\right)\cos\left( \frac{k\pi \eta}{P}\right)  \right].\nonumber
\end{eqnarray}
After a large gauge transformation and making use of eq.(\ref{C0exp}) we find,
\begin{eqnarray}
& &  (C_2- (B_2 +\Delta)C_0)\Big]_{\sigma=\epsilon}^{\sigma=-\epsilon}= f_5- f_7( f_4+\Delta) \Big]_{\sigma=\epsilon}^{\sigma=-\epsilon}=\nonumber\\
& & 8 \sum_{k=1}^\infty a_k \frac{k \pi}{P}\left[ \sin\left( \frac{k\pi \eta}{P}\right)  -\left( \frac{k\pi}{P}\right)\cos\left( \frac{k\pi \eta}{P}\right) (\eta-\frac{9\Delta}{2})  \right].
\label{c2b2c02}
\end{eqnarray}

\section{String embedding in the $\eta$ direction} \label{appendixW}
 In this appendix, we study the embedding of a string which can be extended in $t,x_1$ and $\eta$ directions of our background in the eq.(\ref{tendconfig}) to further study the screening scenario in the dual QFTs. Following the formulation in the subsection \ref{screensec}, for a general string configuration parameterised in terms of $(\tau,\gamma)$ possibly extended in $t,x_1,r$ and $\eta$ 
the  Nambu-Goto action reads,
\begin{eqnarray}
& &S_{NG}= T T_{F1}\int d\gamma \sqrt{F^2 x'^2+ G^2 r'^2+ S^2 \eta'^2},\label{NGgeneric2}\\
& & 
F^2=\left(\frac{2\tilde{g}^2}{9}\right)^2 f_1^2(r,\sigma^*,\eta)H(r) r^4,~~ ~~~~
G^2=\left(\frac{2\tilde{g}^2}{9}\right)^2 f_1^2(r,\sigma^*,\eta)\frac{H(r) r^2}{f(r)},\nonumber\\
& & S^2=\left(\frac{2\tilde{g}}{9}\right) f_1^2(r,\sigma^*,\eta) f_3(r,\sigma^*,\eta) H(r)^{1/2} r^2.\nonumber
\end{eqnarray}
Now we assume embedding at a fixed radial coordinate:
\begin{eqnarray}
& & t=\tau,~~x_1=\gamma,~~r=\bar r,~~\eta=\eta(x_1).
\end{eqnarray}
This action should be minimised to understand if a configuration extending in the $\eta$-direction can reach the closest flavour group. Further generalization will be the generic configuration, which can stretch freely in both $r$ and $\eta$, which we leave for a later study. 

For $r=\bar{r}$ and $\sigma^*=0$, we find
\begin{eqnarray}
& & S_{NG}= T_{F1}\int d\tau d\gamma \sqrt{\det[g_{\alpha \beta}]}, =T_{F1} T \int d\eta \sqrt{F^2 + S^2 \eta'^2}, \label{NGwilson4}
\end{eqnarray}
Using the relations in Appendix \ref{usefulidentities}, one has the simplified results
\begin{eqnarray}
& &     f_1^2(\bar r,\sigma^*,\eta)= \frac{9\pi^2}{4 X^4} \left( \frac{3 X^4 \sigma \partial_\sigma V +\sigma^2\partial_\eta^2 V}{\partial_{\eta}^2  V}\right), \qquad  f_3(\bar r,\sigma^*,\eta)= \frac{X^2\partial_{\eta}^2  V}{3 \sigma \partial_\sigma V} .\\
& & f_1^2 f_3=\frac{3\pi^2}{4 X^2}\left( \frac{\sigma\partial_\eta^2 V}{\partial_\sigma V} + 3 X^4 \right)|_{\sigma^*=0}\simeq \frac{9\pi^2}{4}X^2(\bar{r}),~~~ S^2\simeq \frac{\pi^2}{2} \tilde{g}^2 \bar{r}^2.\nonumber\\
& &     f_1^2(\bar r,\sigma^*,\eta)= \frac{9\pi^2}{4 X^4} \left( \frac{3 X^4 \sigma \partial_\sigma V +\sigma^2\partial_\eta^2 V}{\partial_{\eta}^2  V}\right)|_{\sigma^*=0}\simeq \frac{27\pi^2}{4}\frac{-\hat V}{\partial_{\eta}^2  \hat V},\nonumber\\
& &  F^2 \simeq \frac{27\pi^2}{4} \left(\frac{2\tilde{g}^2}{9}\right)^2\frac{-\hat V(0,\eta)}{\partial_{\eta}^2  \hat V(0,\eta)} H(\bar r) \bar r^4 . \nonumber
\end{eqnarray}

Hence we have (with $\gt=\sqrt{\frac{9}{2}}$)
\begin{equation}
    S_{NG}= T_{F1} T ~\int \sqrt{ \frac{27\pi^2}{4}\frac{-\hat V(0,\eta)}{\partial_{\eta}^2  \hat V(0,\eta)} H(\bar r) \bar r^4+\frac{\pi^2}{2} \tilde{g}^2 \bar{r}^2\eta'^2} ~d\eta.
\end{equation}
We intend to analyse this action for a specific Rank function. We take the example presented in eq.(\ref{ak1}). The rank function corresponding to the quiver is,
\[ {\cal R}(\eta) = \begin{cases} 
N\eta & 0\leq \eta \leq (P-1) \\
N(P-1) (P-\eta)& (P-1)\leq \eta\leq P .
\end{cases}
\]
With the potential $\hat{V}(\sigma,\eta)$ 
\begin{eqnarray}
& & \hat{V} = \frac{N P^3}{2 \pi ^3} \text{Re} \left(\text{Li}_3(-e^{-\frac{\pi}{P}  (| \sigma |+i+i \eta  )})-\text{Li}_3(-e^{-\frac{\pi}{P}  (| \sigma |-i+i \eta )}) \right) \, .
\end{eqnarray}
We will set $N=1, P=10$ to perform the numerical analysis. We assume that the quark-antiquark pair is inserted in the first gauge node corresponding to the boundary condition for the string being in $\eta^*=1$. The string is supposed to go inside the bulk in the $\eta$ direction up to a point $\eta_0$ and return back, forming a U-shape embedding. Following the analysis of eqs.(\ref{QQ separation},\ref{QQ energy}) we have
\begin{align}
V_{eff}\left(  \eta \right) &=\frac{F\left(  \eta \right) }{F\left( \eta_{0}\right)S\left(  \eta
            \right) }\sqrt{F^{2}\left(  \eta \right) -F^{2}\left( \eta_{0}\right)}\ ,\\
        L_{QQ}\left( \eta_{0}\right) &=2\int_{\eta^*}^{\eta_{0} }
        \frac{dz}{V_{eff}(z) }\, , \label{QQseparationet}  \\
    E_{QQ}\left( \eta_{0}\right) &=F\left( \eta_{0}\right) L_{QQ}\left( \eta_{0}\right)
            +2\int_{\eta^*}^{\eta_{0}}dz\frac{S\left( z\right) }{F\left( z\right) }
            \sqrt{F\left( z\right) ^{2}-F\left( \eta_{0}\right) ^{2}}
            \label{QQenergyet} \, .
    \end{align}
The resulting plots for the separation of quark-antiquark pair and their energy are provided in Figure \ref{fig:etaW}. Only $r$ dependence in the energy function appears from $S$ which minimizes in $\bar r=r^*$. There are two $\eta_0$ values for the same separation function $L_{QQ}$. The energy increases by separating the quark and anti-quark pair as the embedded string is going deeper inside the $\eta$ direction, but the solution becomes unstable for the larger $\eta_0$. 
\begin{figure}[h]
    \centering
    \includegraphics[scale=0.7]{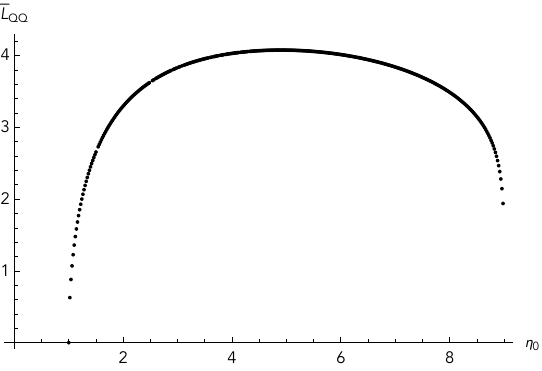} \hspace{1cm}
    \includegraphics[scale=0.7]{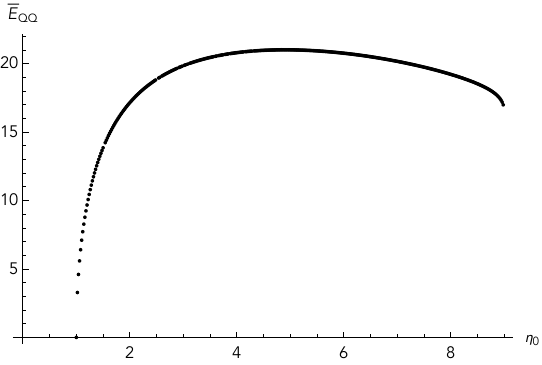}
    \includegraphics[scale=0.7]{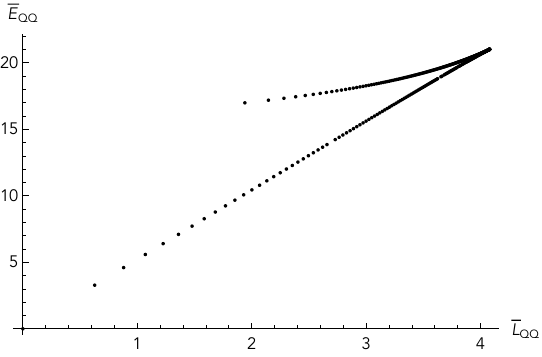}
    \caption{The length of the string between the quark anti-quark pair and the energy with the choice $c=1, \mu=1$.  The length is redefined to be $\bar L_{QQ}=L_{QQ}/(\frac{27\pi^2}{4}H(\bar r) \bar r^4)$ and energy as $\bar E_{QQ}=E_{QQ}/(S(\bar r))$.}
    \label{fig:etaW}
\end{figure}

\end{document}